\documentclass[a4paper,11pt,pdftex]{JHEP3}
\pdfoutput=1
\usepackage{macros} 
\usepackage{amssymb,amsmath}
\usepackage[pdftex]{graphicx}
\usepackage{booktabs,dcolumn}
\newcolumntype{C}{>{$}c<{$}} 
\makeatletter
\def\@xcmidrule{\ifx\@tempa\cmidrule\vskip-\@thisrulewidth
     \global\@lastruleclass=\@ne\else
     \ifx\@tempa\morecmidrules\vskip \cmidrulesep
     \global\@lastruleclass=\@ne\else
     \vskip \belowrulesep\global\@lastruleclass=\z@\fi\fi
     \ifnum0=`{\fi}}
\makeatother
\title{%
Non-perturbative improvement of quark mass renormalization in two-flavour 
lattice QCD
}

\author{%
\begin{flushleft}
\vspace{-0.5cm}
\vbox{\includegraphics{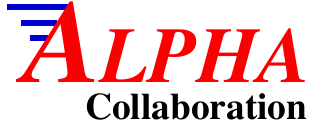}}
\end{flushleft}
}

\author{%
Patrick Fritzsch$^{\,a,b}$,
Jochen Heitger$^{\,b}$ and
Nazario Tantalo$^{\,c,d}$\\

$^{\small a}$
University of Southampton, 
School of Physics and Astronomy\\\hspace{0.5em}
Highfield, Southampton~SO17~1BJ, United Kingdom\\
$^{\small b}$
Westf\"alische Wilhelms-Universit\"at M\"unster,
Institut f\"ur Theoretische Physik\\\hspace{0.5em} 
Wilhelm-Klemm-Stra{\ss}e~9, D-48149 M\"unster, Germany\\
$^{\small c}$
INFN, Sezione di Roma ``Tor Vergata''\\\hspace{0.5em}  
c/o Universit\`{a} di Roma ``Tor Vergata'',
Dipartimento di Fisica,\\\hspace{0.5em} 
Via della Ricerca Scientifica~1, I-00133 Rome, Italy\\
$^{\small d}$
Centro ``Enrico Fermi''\\\hspace{0.5em} 
c/o Compendio Viminale, I-00184 Rome, Italy\\

E-mail: \email{p.fritzsch@soton.ac.uk},\,\email{heitger@uni-muenster.de}\,,\\
\hphantom{E-mail: }\email{nazario.tantalo@roma2.infn.it}
}

\preprint{%
MS-TP-10-06\\
SHEP-1010\\
\today
}

\abstract{%
We non-perturbatively determine the renormalization constant and the 
improvement coefficients relating the renormalized current and subtracted 
quark mass of (quenched) valence quarks propagating in a sea of ${\rm O}(a)$ 
improved two massless quarks.
We employ the Schr\"odinger functional scheme and fix the physical extent 
of the box by working at a constant value of the renormalized coupling.
Our calculation yields results which cover two regions of bare parameter 
space.
One is the weak-coupling region suitable for volumes of about half a fermi.
By making simulations in this region, quarks as heavy as the bottom can be 
propagated with the full relativistic QCD action and renormalization 
problems in HQET can be solved non-perturbatively by a matching to QCD in 
finite volume.
The other region refers to the common parameter range in large-volume
simulations of two-flavour lattice QCD, where our results have particular
relevance for charm physics applications.
}


\begin{document}

\section{Introduction}
\label{Sec_intro}
Hadronic observables can be calculated with non-perturbative accuracy by 
recurring to lattice QCD. Among the shortcomings of all the lattice 
formulations of the continuum theory are cutoff effects, an unavoidable
source of systematic errors that is particularly relevant in the case of 
flavoured quantities.
On the one hand, cutoff effects can be quantified and removed by repeating 
the calculations with different lattice spacings and by extrapolating the 
results to the continuum limit. 
On the other hand, the bigger are the leading cutoff effects the harder will 
be the extrapolations and/or the computational effort required to perform 
simulations at smaller values of the lattice spacing. 

There exist lattice formulations of the continuum theory carrying leading 
cutoff effects proportional to the square of the lattice spacing. 
$\Or(a)$ improvement is achieved at the price of a high 
computational cost by working with lattice Dirac operators, solutions of the 
celebrated Ginsparg-Wilson relation, that preserve at fixed cutoff a lattice 
version of chiral symmetry (see ref.~\cite{reviews:ChLFkaplan} for a recent 
review). 
Automatic $\Or(a)$ improvement of some or all physical observables can be 
also achieved by introducing in the naive lattice fermion action a 
dimension-five operator preserving a non-singlet chiral symmetry while breaking 
at the same time the doubling symmetry and another symmetry of the continuum 
action, e.g. parity in the case of the twisted mass formulation~\cite{impr:tmQCD}. 

In this paper we concentrate on the Wilson formulation of lattice QCD that 
is very well understood, provides a strictly local fermion action, preserves 
parity and all the vector symmetries of the continuum theory and where
$\Or(a)$ improvement is not automatic, but can be obtained by implementing 
Symanzik's program~\cite{impr:sym1,impr:sym2,impr:SW}. 
Furthermore, improvement can be restricted to on-shell quantities and 
correlation functions at physical distances without any 
loss~\cite{impr:onshell}. 

In particular we calculate non-perturbatively with two massless dynamical
flavours of sea quarks the coefficients $\bm$ and $\ba-\bp$ and the renormalization 
constant $Z$ that are needed in order to renormalize (quenched) valence
quark masses in the improved theory. 
The same quantities have been obtained in the quenched case in 
refs.~\cite{impr:roma2_1,impr:babp,HQET:pap2}, and here we closely follow 
the strategy devised in these papers. We employ the 
Schr\"odinger functional scheme~\cite{SF:LNWW,SF:stefan1,SF:stefan2,impr:lett} 
and fix the physical extent of the box by working at a constant value of the 
renormalized coupling.

Our results span two regions of bare parameter space. 
The first is a weak coupling region corresponding to a fixed physical volume 
of about $0.5\,\fm$. 
Simulations on such a small volume are required in order to study 
heavy-light hadronic systems on the lattice with quarks as heavy as the 
bottom. 
Indeed, given the computing resources available nowadays, b-quarks can be 
propagated with their full relativistic action and without introducing 
unacceptably high cutoff effects only in sufficiently small volumes. 
These simulations can then be used to match non-perturbatively HQET with 
QCD~\cite{HQET:pap1}, and physical results can be obtained by calculating 
in the effective theory the running of the renormalization constants down to 
scales corresponding to sufficiently large physical volumes. 
Alternatively, finite-volume QCD results can be corrected by calculating in 
the full theory the finite-volume effects by means of a finite-size scaling 
technique~\cite{fb:roma2a}. 
These two strategies can be profitably combined as it has been done in the 
quenched theory~\cite{mbfb:Nf0}.

The other region of parameter space of our results spans the common range of 
bare couplings used in large-volume simulations of two-flavour lattice QCD, 
where non-perturbative improvement has particular relevance for charm 
physics applications~\cite{mcbar:RS02,fds:final,lat08:mcharmNf2}.

\section{Improvement conditions at constant physics}
\label{Sec_impr}
The basic idea of on-shell improvement of a lattice field theory such as 
Wilson's formulation of lattice QCD as employed in the present paper 
consists in adding counterterms to the unimproved lattice action and fields 
such that the leading cutoff effects are cancelled in on-shell amplitudes.
Having identified the cutoff effect at fixed $g_0$, where a continuum 
symmetry is broken, the coefficients of the different improvement terms 
need to be fixed by suitable improvement conditions. 
These conditions can either require pure lattice artifacts, i.e.~combinations 
of observables that are known to approach zero in the continuum limit of the 
theory, to vanish, or even more generally the scaling of some renormalized 
quantity.
This then defines the values of the improvement coefficients as a function 
of the lattice spacing $a$ or, equivalently, of the bare gauge coupling, 
$g_0^2=6/\beta$.

In perturbation theory, where in principle lattice artifacts can be obtained 
from any renormalized quantity by subtracting its value in the continuum 
limit, the improvement coefficients are unique.
Beyond perturbation theory, where one wants to determine the improvement 
coefficients by Monte Carlo calculations, they do not remain unique and
depend on the choice of the condition. 
This in turn can be exploited
in order to choose an improvement condition with a good sensitivity to the
coefficient in question.
In Wilson's lattice QCD, the resulting ambiguity is of $\Or(a)$ in the coefficients 
and of $\Or(a^2)$ in the physical observables after improvement. 
Extra care has to be exercised, however, if these ambiguities are large.  
A typical example of such a case, which has already been identified in the 
quenched approximation \cite{impr:babp,HQET:pap2}, is the renormalization of 
the subtracted bare quark mass in the $\Or(a)$ improved theory, which 
amounts to consider QCD with non-degenerate quark masses.

The authors of ref.~\cite{impr:losalamos_3} have discussed the details of
improvement and renormalization of the quark field bilinears in the
presence of mass non-degenerate quarks. According to their results a
flavour non-singlet bilinear $O^{ij}$ renormalizes as
\begin{align}
        O_{\rm R}^{ij} &= Z_O \left[ 1 + b_O\,  \tfrac{1}{2}(a\mqi+a\mqj) + 
                                   \tilde{b}_O\, \nf\, a\mq^{({\rm sea})} \right] O^{ij} \,,
        \label{}
\end{align}
while, because of the presence of disconnected diagrams, flavour singlet
bilinears require a different renormalization constant and additional
improvement coefficients. In the rest of our paper we consider only
flavour non-singlet quantities and work at vanishing sea quark masses so
that we need neither to write down contributions proportional to the
improvement coefficient $\tilde{b}_O$ nor to discuss renormalization
of flavour singlet quantities any further.

Let us illustrate this now in more detail by directly entering the 
discussion of the situation in two-flavour QCD addressed by this work.
To avoid notational overload and since any renormalization scale dependence 
is unimportant for the purpose of this section, we write $\mr$ as shorthand 
for $\mbar(\mu)$ in the following.
\subsection{The PCAC relation and mass non-degenerate quarks}
\label{Sec_impr_pcac}
For the formulation of proper improvement conditions in a non-perturbative
calculation it is generally advantageous to use lattice artifacts that 
derive from a symmetry of the continuum field theory that is not respected 
by the lattice regularization.
A particularly suited choice for the purpose of this work are violations 
of the PCAC relation between the axial vector current and the pseudo-scalar 
density.
Assuming that on-shell $\Or(a)$ improvement has been fully implemented and
using renormalized currents in the lattice regularized theory, it reads
\begin{align}
\evalbig{\, \sdrv\mu(\ar)^a_\mu(x)\ \mathcal{O}\, } 
&=  
2\mr\evalbig{\, (P_\mathrm{R})^a(x)\ \mathcal{O}\, }
\myt \,, &
\sdrv\mu
&=
\half\left(\partial_{\mu}+\partial_{\mu}^{\ast}\right) \myt \,,
\label{eqn:renorm-PCAC-relation}
\end{align}
up to $\Or(a^2)$ corrections, as long as $x$ is not in the support of the
source operator $\mathcal{O}$. 
We adopt the standard renormalization pattern of improved lattice currents 
in a mass independent scheme which is conventionally decomposed as 
\begin{align}
\mr 
&= 
\dfrac{\za(1+\ba a\mq)}{\zp(1+\bp a\mq)}\,m+\Or(a^2) \myt \,, &
m   
&= 
\dfrac{1}{2}\,
\dfrac{\evalbig{\sdrv\mu(\aimpr)^a_\mu(x)\,\mathcal{O}}}
{\evalbig{{P}^a(x)\,\mathcal{O}}} \myt \,.
\label{eqn:PCAC-mass-renorm}
\end{align}
Here, $m$ denotes the bare current quark mass defined through some matrix 
element of the bare PCAC relation, where
$(\aimpr)^a_\mu=A^a_\mu+a\ca\sdrv\mu P^a$ is the $\Or(a)$ improved axial 
current and $\ca(g_0)$ is taken to be the non-perturbative value of the 
two-flavour theory~\cite{impr:ca_nf2}.

On the other hand, starting from the bare quark mass $m_0$ of the 
Lagrangian, we have an alternative representation of the renormalized quark 
mass in terms of the $\Or(a)$ improved subtracted bare quark mass, $\mqtil$, 
viz.
\begin{align}
\mr    
&= 
\zm\,\mqtil \myt \,, &
\mqtil 
&= 
\mq(1+\bm a\mq) \myt \,, &
\mq
&=
m_0-\mc \myt \,.
\label{eqn:subtr-mass-renorm}
\end{align}
By equating eqs.~\eqref{eqn:PCAC-mass-renorm} and 
\eqref{eqn:subtr-mass-renorm} one obtains the bare current quark mass in 
terms of the subtracted bare quark mass:
\begin{align}
m
&= 
\dfrac{\zm\zp}{\za}\,
\dfrac{\left(1+\bm a\mq\right)\left(1+\bp a\mq\right)}
{\left(1+\ba a\mq\right)}\,\mq+\Or(a^2)
\label{eqn:barePCAC-vs-mq} \\[0.5em]
&= 
\dfrac{\zm\zp}{\za}\,\Big[1+\left(\bm+\bp-\ba\right)a\mq
+\Or\big([a\mq]^2\big)\Big]\,\mq+\Or(a^2) \myt \,.
\label{eqn:barePCAC-expand-vs-mq}
\end{align}
The renormalization factor
\begin{align}
Z(\gotildesq) 
&\equiv 
\dfrac{\zm(\gotildesq,a\mu)\,\zp(\gotildesq,a\mu)}
{\za(\gotildesq)}
\label{eqn:def-Z-factor}
\end{align}
is finite, because the divergence of $\zm$ is canceled by that of $\zp$. 
Therefore, $Z$ is only a function of the improved bare coupling, 
$\gotildesq=g_0^2(1+b_{\rm g}a\mq^{({\rm sea})})$, while the second factor 
in~\eqref{eqn:barePCAC-vs-mq} is not. 
Our simulations of the two-flavour theory were performed at vanishing 
sea quark mass ($a\mq^{({\rm sea})}\approx 0$) and the valence quarks entering
the bilinears are treated in the quenched approximation, 
hence $\gotildesq=g_0^2$ holds throughout.%
\footnote{
For a more general treatment without these simplifications, the reader may
consult ref.~\cite{impr:losalamos_3}. 
}
To keep notation short, we do not explicitly write out the 
$g_0^2$-dependence in our improvement coefficients and $Z$ from now on. 

To disentangle $Z$, $\bm$ and $\bp-\ba$ in 
eq.~\eqref{eqn:barePCAC-expand-vs-mq} from each other later on, one has to 
consider non-degenerate current quark masses, $\mri{i}\neq\mri{j}$.
In this theory the renormalized, improved axial currents and densities
are given by a straightforward generalization of what was written down 
earlier \cite{impr:roma2_1}.
More precisely, since the presence of non-degenerate masses breaks flavour 
symmetry, our starting point now are the off-diagonal bilinear fields,
\begin{align}\label{eqn:def-off-diag}
A_{\mu}^{ij}(x) 
&= 
\psibar_i(x)\,\dirac\mu\dirac5\,\psi_j(x) \,, 
&
P_{}^{ij}(x) 
&= 
\psibar_i(x)\,\dirac5\,\psi_j(x)  \,.  
\end{align}
As above, the renormalized PCAC relation
\begin{align}
\evalbig{\, \sdrv\mu (\ar)^{ij}_\mu(x)\ \mathcal{O}\, } 
&=  
(\mri{i}+\mri{j})\,\evalbig{\, (P_\mathrm{R})^{ij}(x)\ \mathcal{O}\,}+\Or(a^2)
\label{eqn:renorm-PCAC-relation-pm}
\end{align}
has to be understood as an operator identity, with
\begin{align}
(\ar)^{ij}(x) 
&= 
\za(\gosq)\left[\,
1+\ba(g_0)\,\half\left(a\mqi+a\mqj\right)
\,\right](A_{\rm I})_{\mu}^{ij}(x) \myt \,, 
\label{eqn:ren-off-diag-A} \\[0.5em]
(\pr)^{ij}(x) 
&= 
\zp(\gosq,a\mu)\left[\,
1+\bp(g_0)\,\half\left(a\mqi+a\mqj\right)
\,\right]P^{ij}(x)
\label{eqn:ren-off-diag-P}
\end{align}
and $(A_{\rm I})_{\mu}^{ij}=A_{\mu}^{ij}+a\ca\sdrv\mu P^{ij}$.
The renormalization constants $\za$ and $\zp$ involved are the same as those 
in the theory with two mass-degenerate quarks, and the $b$-coefficients now 
multiply cutoff effects proportional to the average $\half(a\mqi+a\mqj)$ of 
the subtracted bare quark masses. 

Correspondingly, also the subtracted quark masses of each flavour $i$ get 
separately improved, but share the renormalization factor of the degenerate 
case, viz.
\begin{align}
\mri{i} 
&= 
\zm\,\mqitil \myt \,. &
\mqitil  
&= 
\mqi\,(1+\bm a\mqi) \myt \,, &
\mqi
&=
m_{0,i}-\mc \myt \,,
\end{align}
and in analogy to eq.~\eqref{eqn:barePCAC-vs-mq},
\begin{align}
\mri{i}
&=
\dfrac{\za}{\zp}\,Z\,\mqi\,(1+\bm a\mqi) \myt \,, &
Z
&= 
\dfrac{\zm\zp}{\za} \myt \,.
\label{eqn:renormPCAC-vs-mq}
\end{align}
By inserting the last equation into the PCAC relation
\eqref{eqn:renorm-PCAC-relation-pm}, and by solving for the associated bare 
current quark masses, we arrive at the generalization of 
eqs.~\eqref{eqn:barePCAC-vs-mq} and~\eqref{eqn:barePCAC-expand-vs-mq} for
the case of mass non-degenerate quarks:
\begin{align} 
        \lefteqn{\dfrac{\evalbig{ \big(\sdrv{0}A_{0}^{ij}+a\ca\drvstar{0}\drv{0}P^{ij}\big)\mathcal{O}\, } }{2\,\evalbig{ P^{ij}\,\mathcal{O} } }=
}\label{eqn:barePCAC-pm-expand-vs-mq}\\[0.5em] 
&\qquad      
Z\left[\,\half\left(\mqi+\mqj\right)+\half\,\bm\left(a\mqi^2+a\mqj^2\right) 
-\quart\left(\ba-\bP\right)a\left(\mqi+\mqj\right)^2\,\right] \myt \,,\notag 
\end{align}
again up to terms of $\Or(a^2)$. 
Moreover, we introduce a third renormalized mass defined as the mean value 
of two given ones,
\begin{align}
\half\,(\mri{1}+\mri{2}) 
&=
\zm\left[\,
\half\,(\mqn{1}+\mqn{2})+\bm\,a\,\half\,(\mqn{1}^2+\mqn{2}^2)
\,\right] \myt \,,
\label{eqn:m1+m2-renorm}
\end{align}
which differs from the one obtained by the renormalization of the mean 
$\moi{3}=\frac{1}{2}(\moi{1}+\moi{2})$ taken at the level of bare quark 
masses,
\begin{align}
\mri{3}\equiv\zm\,\mqn{3}(1+\bm a\mqn{3}) 
&=  
\zm\left[\,
\half\,(\mqn{1}+\mqn{2})+\bm\,a\,\quart\,(\mqn{1}+\mqn{2})^2
\,\right] \myt \,.
\label{eqn:m3-renorm}
\end{align}
The difference of the quadratic terms in eq.~\eqref{eqn:m1+m2-renorm}
and~\eqref{eqn:m3-renorm} can be used to estimate $\bm$. 
By building appropriate differences with non-degenerate quarks it is 
possible to construct estimators for the improvement coefficient $\bm$ and 
$\ba-\bp$ separately.

In the following we show how one can extract $\ba-\bp$, $\bm$ and $Z$ by 
applying eq.~\eqref{eqn:barePCAC-pm-expand-vs-mq} with a few external 
operators resp.~kinematical conditions~\cite{impr:roma2_1,impr:babp}
(see also~\cite{impr:losalamos_1}).
\subsection{Schr\"odinger functional correlation functions}
\label{Sec_impr_sfcfs}
Particularly in the context of the QCD Schr\"odinger functional (SF)
and its lattice formulation \cite{SF:LNWW,SF:stefan1,SF:stefan2,impr:lett}, 
the use of PCAC has proven to 
be very advantageous to solve improvement and renormalization problems in 
many situations, because there one has a large flexibility to choose 
appropriate improvement and renormalization conditions.
Beyond that, one can compute, e.g., the improvement coefficients also for 
small values of the bare coupling $g_0$, thereby making contact with their 
perturbative expansions. 
An overview is given, for instance, in~\Ref{reviews:NPRrainer_nara}.

If $\zeta$ and $\zetabar$ denote 
boundary quark and anti-quark fields at Euclidean time $x_0=0$
\cite{impr:pap1}, the field products summed over the spatial volume,
\begin{align}        
\obi 
&= 
\dfrac{a^6}{L^3}\sum_{\vecu,\vecv}  
\zetabar(\vecu)\,\dirac5\frac{1}{2}\,\tau^a\,\zeta(\vecv) \myt \,, &
\obf 
&= 
\dfrac{a^6}{L^3}\sum_{\vecu,\vecv}  
\zetabarprime(\vecu)\,\dirac5\frac{1}{2}\,\tau^a\,\zetaprime(\vecv) \myt \,, 
\label{eqn:PS-ops}
\end{align}
yield pseudo-scalar boundary sources projected onto zero momentum.
From these, the boundary-to-bulk resp. forward SF correlation functions in 
the pseudo-scalar channel are constructed from the axial current and density
as 
\begin{align}
\fa(x_0) 
&= 
-\dfrac{a^3}{2}\sum_{\vecx}
\big\langle A^a_0(x)\,\obi\big\rangle \myt \,, &
\fp(x_0) 
&= 
-\dfrac{a^3}{2}\sum_{\vecx}
\big\langle P^a(x)\,\obi\big\rangle \myt \,. 
\label{eqn:def-fa-fp}
\end{align}
The corresponding bulk-to-boundary resp. backward correlation functions, $g_{\rm A, P}$,
are related to the former by a time reflection and fall off with $T-x_0$ from their 
sources at $x_0=T$ as do the forward correlation functions with $x_0$ from $x_0=0$. 
In a vanishing background field, this time reflection symmetry can be utilized by
taking their mean before building secondary quantities, in order to reduce the statistical noise.
In the theory with mass non-degenerate quarks the bare correlation functions generalize 
to
\begin{align}
\fa^{ij}(x_0) 
&=
-\dfrac{a^3}{2}\sum_{\vecx}\big\langle A_0^{ij}(x)\,\ob^{ji}\,\big\rangle \myt \,, &
\fp^{ij}(x_0) 
&=
-\dfrac{a^3}{2}\sum_{\vecx}\,\big\langle P^{ij}(x)\,\ob^{ji}\,\big\rangle \myt \,.
\label{eqn:def-fA-fP-ij}
\end{align}
We do not apply a summation convention here, and the boundary sources $\ob^{ji}$ 
are formally composed by non-degenerate boundary quark and anti-quark fields
in the same way as the currents in eq.~\eqref{eqn:def-off-diag}. 
Backward correlation functions $g^{ij}_{{\rm A},{\rm P}}(T-x_0)$ are defined 
analogously. 

Including the $\Or(a)$ correction of the axial current as in 
eq.~\eqref{eqn:ren-off-diag-A}, the improved bare current quark 
masses\footnote{
This expression for the PCAC masses is only $\Or(a)$ improved up to a factor
$1+\half(\ba-\bP)(a\mqi+a\mqj)$ for quark mass dependent cutoff effects. 
}
expressed in terms of the just introduced SF correlation functions are given 
by
\begin{align}
m_{ij}(x_0)\equiv m_{ij}(x_0;L/a,T/L,\theta)
&= 
\dfrac{\sdrv0\fa^{ij}(x_0)+a\ca\drvstar{0}\drv{0}\fp^{ij}(x_0)}
{2\,\fp^{ij}(x_0)} \myt \,,
\label{eqn:mpcac_x0-ij}
\end{align}
where, only here, we explicitly indicate their additional dependence on 
$L/a$, $T/L$ and the periodicity angle $\theta$ in the boundary conditions
of the fermion fields.
In the degenerate case ($i=j$), $m_{ij}$ reduces to the non-singlet PCAC mass of a 
flavour degenerate doublet. 
In this definition, $\sdrv0$~denotes the average of the usual forward and 
backward derivatives but, as in \cite{impr:roma2_1,impr:babp,HQET:pap2}, 
we have also employed current quark masses involving improved derivatives, 
obtained by the replacements
\begin{align}
\sdrv0
&\rightarrow
\sdrv0\left(1-{\T \frac{1}{6}}\,a^2\drvstar{0}\drv{0}\right) \myt \,, &
\drvstar{0}\drv{0}
&\rightarrow
\drvstar{0}\drv{0}
\left(1-{\T \frac{1}{12}}\,a^2\drvstar{0}\drv{0}\right) \myt \,,
\label{eqn:deriv}
\end{align}
in order to eventually estimate $\bm$, $\ba-\bp$ and $Z$. 
When acting on smooth functions, these derivatives have an error of 
$\Or\big({g_0^2a^2},a^4\big)$. 
\subsection{Estimators for $\ba-\bP$, $\bm$ and $Z$}
\label{Sec_impr_RX}
Following the strategy already proposed for the quenched computation 
in \cite{impr:babp,HQET:pap2}, we form ratios of suitable combinations of 
degenerate and non-degenerate current quark masses.
Representing eq.~\eqref{eqn:barePCAC-pm-expand-vs-mq} through the $m_{ij}$
now, it turns into
\begin{align} 
m_{ij} 
&=      
Z\left[\,\half\left(\mqi+\mqj\right)+\half\,\bm\left(a\mqi^2+a\mqj^2\right) 
-\quart\left(\ba-\bP\right)a\left(\mqi+\mqj\right)^2\,\right] \,.
\label{eqn:barePCAC-ij-expand-vs-mq}
\end{align}
This enables us to derive direct estimators for $\ba-\bP$, $\bm$ and $Z$
in terms of various bare PCAC quark masses $m_{ij}$, which can by computed 
via SF correlation functions according to eq.~\eqref{eqn:mpcac_x0-ij}.
By exploiting the symmetry properties~\cite{impr:babp} of 
eq.~\eqref{eqn:barePCAC-ij-expand-vs-mq} and keeping only the relevant terms, 
i.e. 
\begin{align}
2a\mij{12}-(a\mij{11}+a\mij{22}) 
    &=
    Z\,\half\,(\ba-\bp)\left(a\mqn{1}-\mqn{2}\right)^2\left[\, 1 + \Or\big(a\mqn{1}\!+\!a\mqn{2}\big)  \right] \,,
    \label{eqn:expand-for-RAP} \\[0.4em]
a\mij{11}-a\mij{22} 
    &=
    Z\left(a\mqn{1}-a\mqn{2}\right)\left[\, 1 +\Or\big(a\mqn{1}\!+\!a\mqn{2}\big)\right] \,,
\label{eqn:expand-for-Rm} \\[0.4em]
a\mij{12}-a\mij{33} 
    &=
    \quart\,\bm\,Z\,\left(a\mqn{1}-a\mqn{2}\right)^2\left[\,1+\Or\big(a\mqn{1}\!+\!a\mqn{2}\big)\,\right] \myt \,,
    \label{eqn:expand-for-RZ}
\end{align}
it is straightforward to deduce the following ratios, 
which define estimators for the desired improvement coefficients and
renormalization constant:
\begin{align}
\RAP 
&\equiv 
\dfrac{2\left(2\mij{12}-\mij{11}-\mij{22}\right)}
{\left(\mij{11}-\mij{22}\right)\left(a\mqn{1}-a\mqn{2}\right)} = 
(\ba-\bP)\left[ 1+\Or\big(a\mqn{1}+a\mqn{2}\big)\right] \myt \,,
\label{eqn:estim-RAP} \\[0.4em]
\Rm 
&\equiv 
\dfrac{4\left(\mij{12}-\mij{33}\right)}
{\left(\mij{11}-\mij{22}\right)\left(a\mqn{1}-a\mqn{2}\right)} =
\hphantom{ (\ba- )}\;\bm \left[ 1+\Or\big(a\mqn{1}+a\mqn{2}\big)\right] \myt \,,
\label{eqn:estim-Rm} \\[0.4em]
R_Z 
&\equiv 
\dfrac{\mij{11}-\mij{22}}{\mqn{1}-\mqn{2}}
+\left(\RAP-\Rm\right)\left(a\mij{11}+a\mij{22}\right) = Z + \Or\big(a^2\big)    \myt \,.
\label{eqn:estim-RZ}
\end{align}
While the leading cutoff effects on the estimators for the improvement coefficients
$\RAP$ and $\Rm$ depend (linearly) on the bare subtracted quark masses, the leading
correction to the estimator $R_Z$ for the renormalization factor is $\Or\big(a^2\big)$
after inserting
\begin{align}
\RAP\!-\!\Rm 
&\equiv 
\dfrac{2\left(2\mij{33}-\mij{11}-\mij{22}\right)}
{\left(\mij{11}-\mij{22}\right)\left(a\mqn{1}-a\mqn{2}\right)}  
= 
\left(\ba\!-\!\bp\!-\!\bm\right)+\Or\big(a\mqn{1}+a\mqn{2}\big) \,.
\label{eqn:RAPn-expanded}
\end{align}
The difference $\RAP-\Rm$ does not depend on the off-diagonal 
mass $\mij{12}$ and the cutoff-dependence of $R_Z$ is the one expected 
for a renormalization constant in the $\Or(a)$ improved theory. Note that the essential
piece for the disentanglement of $\RAP$ and $\Rm$ is the introduction of a 
third mass in accordance with
\begin{align}   
m_{0,3}
&=
\half\,(m_{0,1}+m_{0,2}) && \Leftrightarrow &
\mqn{3}
&=
\half\,(\mqn{1}+\mqn{2}) \myt \,, 
\label{eqn:third-mass}
\end{align}
as already anticipated in eqs.~\eqref{eqn:m1+m2-renorm} 
and \eqref{eqn:m3-renorm} before.

Since the bare current quark masses $m_{ij}$ are functions of the time
coordinate $x_0$, where the operator insertion of the currents $A_0$ and $P$ 
takes place, so do the estimators $R_X$, $X={\rm AP},{\rm m},Z$, depend 
on $x_0$. 
Therefore, one has to specify the exact timeslice range used to compute the 
estimators $R_X$. 
This choice must be considered as being part of a particular improvement 
condition, which uniquely defines the ratios $R_X$ as smooth functions of 
$g_0$.
Within the SF formalism, a good plateau in the local PCAC masses usually 
sets in around the midpoint of the time extension, $T/2$. 
This need not necessarily be the case for observables derived from these 
masses, especially when investigating improvement coefficients which 
genuinely emerge only at order $a$. 
Fluctuations are in general expected to show up, depending on the quantity 
and the region in bare parameter space where the simulation takes place,
but this does not invalidate the once chosen improvement condition.

As we will use $T=3L/2$ later on, our final estimators are defined as 
timeslice averages over the central third of the time extension:
\begin{align}
R_X 
&\equiv
\dfrac{a}{T-L/2+a}\sum_{x_0=L/2a}^{(T-L/2)/a}R_X(x_0) \myt \,, &  
X 
&\in   
\{{\rm AP},{\rm m},Z\} \myt \,.
\label{eqn:RX-average}
\end{align}
Note that the main reason for this is rather to be less dependent on a 
special timeslice --- which means to get a smoother and more reliable 
signal --- than simply decreasing the size of the error.
Furthermore, since the number of timeslices used for the plateau average 
scales with the lattice size, the physical size of this plateau is kept 
constant and hence automatically respects our requirement of a constant 
physics condition to be used below.
\ subsection{General remarks on  improvement conditions at constant physics}
\label{Sec_impr_iccp}
Before proceeding with the actual computation of $\ba-\bP$, $\bm$ and $Z$,
we would like to discuss the virtues of improvement conditions at constant 
physics on more general grounds~\cite{reviews:NPRrainer_nara}.

In the aforementioned quenched investigations \cite{impr:babp,HQET:pap2} it
was found that, e.g., the result for $\ba-\bp$ at $a\approx0.1\,\Fm$ does 
depend significantly on the computational details, namely the timeslice 
location $x_0$ and the exact lattice representation of derivatives. 
(See also the upper right panel in figure~\ref{fig:RX-plateaus} of 
the present work.)
Obviously, this is quite an extreme example of an improvement coefficient 
that is difficult to determine, and its ambiguity from choosing a different
improvement condition may even grow to the same order as the coefficient
itself.
The reason for such a behaviour lies in the fact that the $\Or(a)$ effects 
are not dominating over the $\Or(a^2)$ ones in the considered correlation 
functions. 
Presumably, the $\Or(a)$ effects are just not very large.  

One of the benefits of the Schr\"odinger functional pointed out before is
that improvement conditions formulated through SF correlation
functions can also be studied in perturbation theory. 
Choosing kinematical variables such as $x_0/L$, $T/L$ and $\theta$ exactly 
as in the non-perturbative setup, the expansion of an improvement 
coefficient, here denoted generically by $b$, is calculated as a series
\begin{align}
b(g_0,a/L) 
&\sim 
b^{(0)}(a/L)+g_0^2\,b^{(1)}(a/L)+\Or(g_0^4) \myt \,. 
\end{align}
For our example of $\ba-\bp$ it was observed in \Ref{impr:babp} that the 
perturbative coefficients $b^{(0)}$ and $b^{(1)}$ show a similar dependence 
on the kinematics as the non-perturbative results. 
Our statement on the uniqueness of the improvement coefficients in
perturbation theory at the beginning of this section, however, means that
although generically the functions $b^{(0)}(a/L)$ and $b^{(1)}(a/L)$ will
depend on the kinematical choices made in the improvement condition, 
the values $b^{(0)}(0)$ and $b^{(1)}(0)$ do not and are unique. 

Hence, the following generic picture emerges. 
Once improvement conditions at \emph{constant physics} are imposed, by which
we mean that all length scales in the correlation functions that define the 
improvement condition are kept fixed in physical units, only the lattice 
spacing $a$ changes when $g_0$ is varied.
This is just the situation, to which Symanzik's local effective theory of
cutoff effects can be applied. 
As a consequence, improvement coefficients extracted from different
constant physics conditions are then expected to rapidly approach an 
almost unique function of $g_0$ as $g_0\to 0$.  
While it is clear from this discussion that fixed $L$ in physical 
units is to be preferred whenever possible, in certain situations one may 
want to implement improvement conditions at fixed $L/a$ for practical 
reasons. But in such a setting, the $\Or(a)$ ambiguities in the improvement 
coefficients would not go to zero as $g_0\to 0$ $(a\to 0)$.%
\footnote{%
E.g., improvement conditions at fixed $L/a$ were employed for the
non-perturbative determination of $\csw$ and $\ca$ in quenched 
QCD~\cite{impr:pap3}. 
In this case, however, the small tree-level $a$-effect was subtracted from 
the non-perturbative ones to insure that the improvement coefficients go 
to their tree-level values exactly; furthermore, the conditions were chosen 
such that also $|b^{(1)}(a/L)-b^{(1)}(0)|$ is very small.
} 

In the present situation, where we have to work at finite quark masses, also 
the combination $L\mri{i}$ should be kept constant.
Note, however, that these \emph{constant physics conditions} do not have to 
be satisfied very precisely, because this would affect corrections to an 
$\Or(a)$ term only.
In addition to the studies~\cite{impr:babp,HQET:pap2} in quenched QCD,
such conditions have meanwhile also been imposed for the determination of 
$\ca$~\cite{impr:ca_nf2} and $\za$ \cite{impr:za_nf2_2} in the dynamical 
two-flavour theory.

Let us summarize the discussion of this subsection.
While an unfortunately chosen improvement condition, but implemented at 
constant physics, may even enlarge the cutoff effects for intermediate $a$, 
it guarantees that linear $a$-effects are absent. 
Putting the improvement coefficient to zero or some perturbative 
approximation does \emph{not} guarantee the latter, and the linear 
$a$-effects should at least be estimated in some way.

\section{Numerical computation and results}
\label{Sec_res}
The goal of this section is to non-perturbatively compute the improvement 
factors $\bm(g_0)$ and $[\ba-\bp](g_0)$ as well as the renormalization 
constant $Z(g_0)$ in the $\nf=2$ dynamical case along the lines of the
foregoing section.
\subsection{The constant physics condition}
\label{Sec_res_constphys}
In view of the freedom of possible legitimate choices for kinematical
parameters within the SF formulation of lattice QCD \cite{SF:LNWW}, it was 
found to be advantageous to choose slightly different, particularly adapted 
settings in specifying a suitable constant physics condition for each of the 
parameter regions mentioned in the introduction.

In the SF, the renormalization scale $\mu=1/L$ and thus the physical volume 
are implicitly defined by demanding the non-perturbatively renormalized 
coupling to take a certain value. 
Therefore, to impose our improvement/normalization conditions on a line of
constant physics, it is required to keep all length scales fixed as $g_0$
is varied, which we realize for the purposes of this work as
\bit
\item 
condition~1:
\begin{align}
\theta   
&= 0.5 \myt \,, & 
\gbsq(L) &= 2.989 && \Leftrightarrow &  
\text{$L\approx 0.25\,\Fm$ \,for\, $6.2\lesssim\beta\lesssim 6.6$}
\label{eqn:cond-coupl-2.989}
\end{align}
\item 
condition~2:
\begin{align}
\theta
&= 0 \myt \,, &
\text{$L\approx 1.8\Lstar$ s.th. $\gbsq(\Lstar)=5.5$} 
&& \Leftrightarrow &&  
\text{$L\approx 1.2\,\Fm$ \,for\, $5.2\le\beta\le 5.7$}
\label{eqn:cond-coupl-5.5}
\end{align}
\eit
supplemented by choosing as fixed kinematical parameters
\begin{align}
C_k &= C^\prime_k=0 \; \text{(vanishing background field)} \myt \,, & 
T/L &= 3/2 \myt \,,  
\label{eqn:kin-params-SF-bX-Z}
\end{align}
where $C_k$ and $C_k^\prime$ induce the Dirichlet boundary conditions of the 
gauge field at $x_0=0$ and $T$, respectively.

The first $\beta$-range (`condition~1') refers to a region of weak 
couplings encountered in solving renormalization problems of lattice HQET 
and in studies of B-physics observables in the continuum limit of
finite-volume two-flavour lattice QCD, 
see \Refs{lat07:hqetNf2,lat08:hqettests} and section~\ref{Sec_lcps}.
Our second $\beta$-range (`condition~2') corresponds to lattice spacings of
$0.09\,\Fm\gtrsim a\gtrsim 0.04\,\Fm$~\cite{cern:Nf2light_1,lat07:rainer},
which is the typical coupling region employed in $\nf=2$ large-volume 
simulations.
In this case, the choice $\theta=0$ (periodic boundary conditions) together 
with an intermediate physical volume of $L\approx 1.2\,\Fm$ is guided by 
experiences made in the context of the non-perturbative redetermination of 
the axial current renormalization constant $\za$ in~\Ref{impr:za_nf2_2}.
There it was observed that this is a kinematical region, 
where $a^2$-effects (which can be significant at $a\approx 0.1\,\Fm$) are
better suppressed, implying reduced intrinsic $a^2$-ambiguities in $\za$, 
and SF simulations can be done close to or in the chiral limit of vanishing 
sea quark mass.
The chiral limit can be reached in the simulations, when the infrared cutoff
on the spectrum of the Dirac operator (in the SF dominantly controlled by 
$1/T$) is sufficiently large, so that $T=3L/2\approx 1.8\,\Fm$ is a feasible 
choice for this parameter region and the underlying $\nf=2$ gauge field 
configurations could be shared among the two projects. 

For the generation of our dynamical gauge field configurations with $\nf=2$
massless sea quarks, we employed the HMC algorithm with two pseudo-fermion 
fields as proposed in~\cite{hmc:hasenb1,hmc:hasenb2}.
The particular implementation has been discussed and tested in 
\Refs{Nf2SF:algo,Nf2SF:autocorr} and following the latter, for the 
$\beta$-range and volume of~\eqref{eqn:cond-coupl-5.5} we always chose a 
molecular dynamics trajectory length of $\tau=2$, the only exception being 
$\beta=5.2$, where we set $\tau=1$.
For the $\beta$-range and volume of ~\eqref{eqn:cond-coupl-2.989}, we had 
$\tau=1$ throughout.
To prepare a light sea quark doublet of mass zero in the simulations, the 
corresponding hopping parameter, $\kapl$, was tuned to its critical value,
$\kapc$.

In order to complete the definition of the lines of constant phy\-sics, 
values for the bare PCAC mas\-ses of the valence quarks must be selected. 
In order to get an impression of the quark mass dependence in physical
observables, we consider in analogy to the quenched study~\cite{HQET:pap2} 
two sets,
\begin{align}
\text{set~1}
&: &
L m_{11} &\approx 0   \myt \,, &
L m_{22} &\approx 0.5 \myt \,,
\label{eqn:set1} \\
\text{set~2}
&: &
L m_{11} &\approx 0   \myt \,, &
L m_{22} &\approx 2.5 \myt \,,
\label{eqn:set2}
\end{align}
where throughout this section all bare current quark masses are understood
to be calculated according to an averaging prescription analogous to
eq.~\eqref{eqn:RX-average}.
The appearing local current quark masses $\mij{ii}(x_0)$ are evaluated using 
the standard lattice derivative as implied, following standard conventions, 
by its definition \eqref{eqn:mpcac_x0-ij} based on the $\Or(a)$ improved 
PCAC relation \eqref{eqn:renorm-PCAC-relation-pm}.

The choice on $Lm_{22}$ in set~1 is motivated by experiences from the 
quenched investigation, \Ref{impr:babp}, where it was argued to be 
advantageous w.r.t.~the size of $\Or(a)$ ambiguities encountered in 
perturbation theory.
By contrast, the choice of $L m_{22}\approx2.5$ in set~2 is closer to the 
typical $\rm b$-quark region itself. The conditions on $Lm_{22}$
have been satisfied by adjusting the hopping parameters $\kappa_{2}$, 
responsible for the mass value of the heavy valence quark flavour.
The hopping parameter $\kappa_{3}$ needed to extract $\bm$,
which fixes a third mass value according to eq.~\eqref{eqn:third-mass},
is given by $\kappa_{3}={2\kappa_{1}\kappa_{2}}\big/{(\kappa_{1}+\kappa_{2})}$
and was directly implemented in the measurement code to avoid rounding 
errors.
Furthermore, we exploit the additional time-reflection symmetry of the 
forward and backward boundary-to-bulk fermionic SF correlation functions 
introduced in the previous section by summing them in the analysis
accordingly.
\subsection{Results in the weak-coupling region of finite-volume QCD}
\label{Sec_res_RXweak}
%
\TABLE[t]{
\centering
\small
\renewcommand{\arraystretch}{1.25}
\begin{tabular}{CCCCCCCCC}
\toprule
L/a & T/a & \beta  &&                    & \kappa_{i}  && Lm_{ii}          \\\midrule
 10 & 15  & 6.1906 && \kappa_{1}         &  0.1360160  && -0.0054(7)\hpo   \\[-0.25em]   
    &     &        && \kappa_{2}         &  0.1343182  && +0.5004(6)\hpo   \\[-0.25em]
    &     &        && \widehat\kappa_{2} &  0.1276218  && +2.5004(7)\hpo   \\
 12 & 18  & 6.3158 && \kappa_{1}         &  0.1357930  && -0.0062(10)      \\[-0.25em]
    &     &        && \kappa_{2}         &  0.1343782  && +0.5008(9)\hpo   \\[-0.25em]
    &     &        && \widehat\kappa_{2} &  0.1287549  && +2.5007(10)      \\
 16 & 24  & 6.5113 && \kappa_{1}         &  0.1354410  && -0.0101(13)      \\[-0.25em]  
    &     &        && \kappa_{2}         &  0.1343872  && +0.4949(13)      \\[-0.25em]
    &     &        && \widehat\kappa_{2} &  0.1301456  && +2.4955(13)      \\
 20 & 30  & 6.6380 && \kappa_{1}         &  0.1351630  && +0.0176(9)\hpo   \\[-0.25em]
    &     &        && \kappa_{2}         &  0.1343562  && +0.5024(9)\hpo   \\[-0.25em]
    &     &        && \widehat\kappa_{2} &  0.1309650  && +2.5039(10)      \\
\bottomrule
\end{tabular}
\caption{%
Results after tuning the hopping parameters in our small-volume computations
to different values of the dimensionless (and plateau-averaged) PCAC masses 
$Lm_{11}$ and $Lm_{22}$ as specified in eqs.~\eqref{eqn:set1} 
and~\eqref{eqn:set2}. 
Set~1 and set~2 share the same sea quark content, given by 
$\kapl=\kappa_c=\kappa_1$. 
The hopping parameters marked by a hat belong to set~2.
}\label{tab:tune-LmPCAC}

}
%
\FIGURE[t]{
\includegraphics[width=\textwidth]{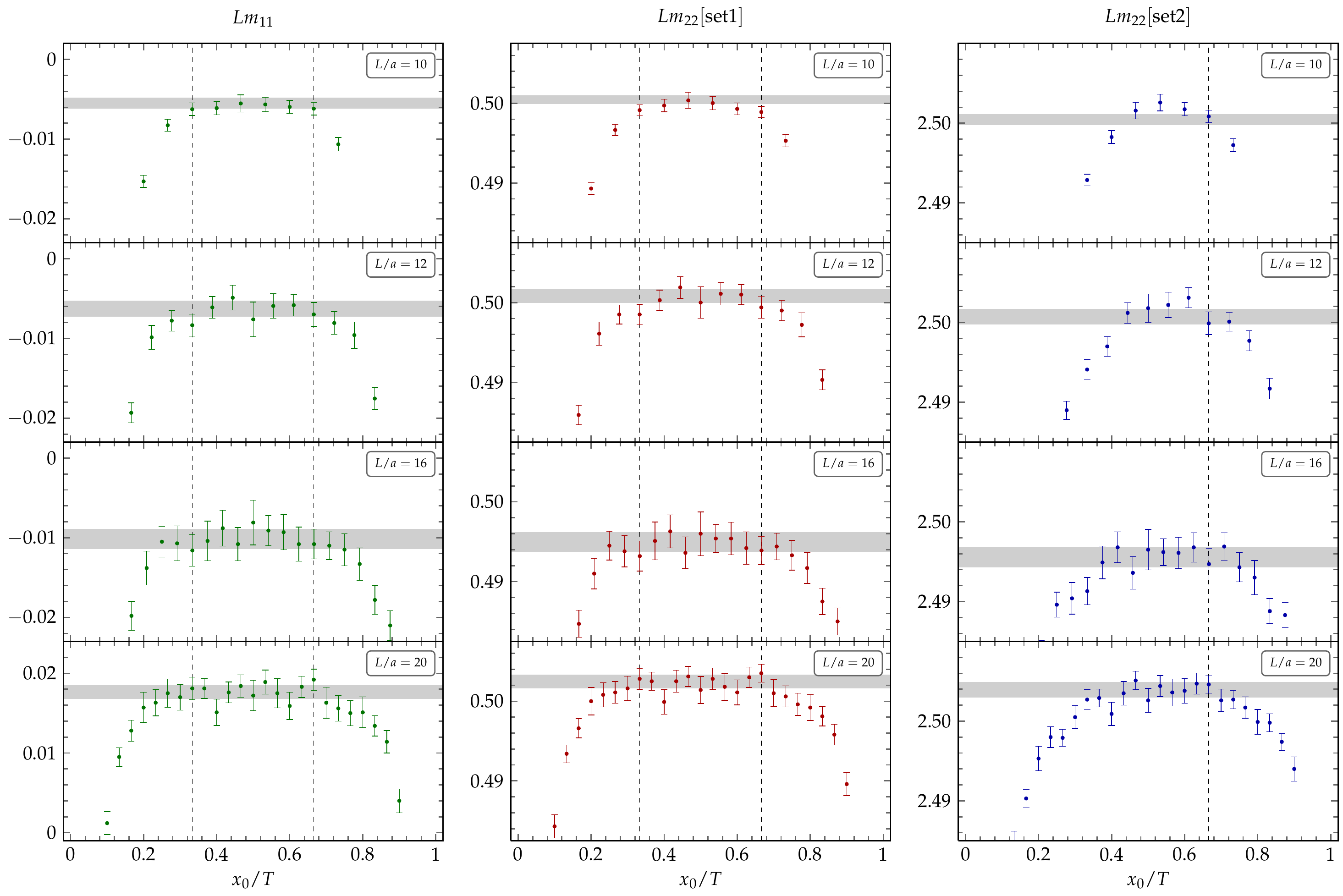}
\vspace{-0.875cm}
\caption{%
Time dependence of dimensionless PCAC masses obtained in small-volume QCD 
simulation after tuning the hopping parameters of the valence quarks as 
given in table~\ref{tab:tune-LmPCAC}.
The dashed lines define the plateau region used for the averaging as
described in the text. 
Corresponding averaged mass values are shown as error bands in each subplot. 
$Lm_{11}\approx 0$, and $Lm_{22}\approx 0.5$ for set~1 resp. 
$Lm_{22}\approx 2.5$ for set~2 are fulfilled within the desired accuracy. 
}\label{fig:avgd-L0mii}
}
%
The parameter values for calculating the ratios $R_X$ in $L\approx0.25\,\fm$ 
are given in table~\ref{tab:tune-LmPCAC}.
Our statistics consists of $200-300$ gauge field configurations
stored at frequencies $5-10$ in units of molecular dynamics trajectories.
The time extent $T=3L/2$ is chosen such that the plateau average,
eq.~\eqref{eqn:RX-average}, is less affected by finite-size effects stemming 
from the boundaries compared to the other natural choice $T=L$.
After having produced gauge field configurations with $\nf=2$ massless 
dynamical quarks, we performed one or two steps of tuning to properly adjust 
$\kappa_{2}$ and satisfy the conditions\footnote{
Similar to the situation in~\cite{impr:babp,HQET:pap2}, this is to a
sufficient precision equivalent to keeping the corresponding renormalized 
masses $L\za m/\zp$ fixed, as for the considered couplings the entering 
overall renormalization constant barely varies.
} 
imposed on $Lm_{22}$ (see also figure~\ref{fig:avgd-L0mii}).
The estimators $R_X$, $X\in\{{\rm AP},{\rm m},{Z}\}$, are defined in terms 
of PCAC masses, which are secondary quantities themselves, built from SF 
correlation functions by means of eq.~\eqref{eqn:mpcac_x0-ij}. 
For the evaluation of all our observables and their statistical errors we
hence employed the $\Gamma$-method \cite{MCerr:ulli}, which directly 
analyzes their autocorrelation functions.
As a cross-check, a standard Jackknife procedure was applied as well.
The results for $R_X$ from improvement condition~1 for set~1 and set~2 
are listed in table~\ref{tab:RX-direct-results}.
%
\TABLE[t]{
\centering
\small
\renewcommand{\arraystretch}{1.25}
\begin{tabular}{c*{6}{C}} 
\toprule
 set & L/a & \beta   &  \ba\!-\!\bp  &  \bm            &  Z            &  \ba\!-\!\bp\!-\!\bm \\\midrule
  1  & 10  & 6.1906  & +0.0027(11)   &  -0.6576(15)    &  +1.10418(19) &  +0.6637(8)\hpo      \\
     & 12  & 6.3158  & +0.0011(37)   &  -0.6666(26)    &  +1.10522(29) &  +0.6653(13)         \\
     & 16  & 6.5113  & -0.006(11)\hpo& -0.6637(36)     &  +1.10395(29) &  +0.6614(22)         \\
     & 20  & 6.6380  & -0.0050(29)   &  -0.6636(48)    &  +1.10333(23) &  +0.6648(21)         \\\midrule
  2  & 10  & 6.1906  & +0.07261(41)  &  -0.56417(38)   &  +1.09522(11) &  +0.63820(37)        \\
     & 12  & 6.3158  & +0.05177(61)  &  -0.57800(66)   &  +1.09747(13) &  +0.62971(60)        \\
     & 16  & 6.5113  & +0.02950(60)  &  -0.5955(10)\hpo& +1.09945(13)  &  +0.62617(69)        \\
     & 20  & 6.6380  & +0.02101(68)  &  -0.6077(11)\hpo&  +1.09997(11) &  +0.62914(71)        \\
\bottomrule
\end{tabular} 
\caption{%
Summary of results on the improvement coefficients and the renormalization 
factor under consideration in the volume with $L\approx0.25\,\Fm$, for both 
sets of heavy quark masses (resp. constant physics conditions) using 
plateau-averaged masses and improved derivatives.
}\label{tab:RX-direct-results}

}
%

The two plots on the r.h.s. of figure~\ref{fig:RX-plateaus} show the 
timeslice dependence $\RAP(x_0)$ and $\Rm(x_0)$ for set~1 at $L/a=12$ and
$\beta=6.3158$. 
The averaged estimators that represent our main results are shown as error 
bands. 
The filled symbols refer to improved time derivatives, 
\eqref{eqn:deriv}, while the open symbols are obtained using the standard 
derivatives. 
Their difference can become quite large especially for $\RAP$, as can be 
inferred from the plot. 
This is expected to be a pure cutoff effect and underlines the necessity to 
use improved derivatives. 
A plateau in $\RAP(x_0)$ is visible in the data for both, standard and 
improved derivatives, but it is more pronounced for the latter.

\FIGURE[t]{
\includegraphics[width=\textwidth]{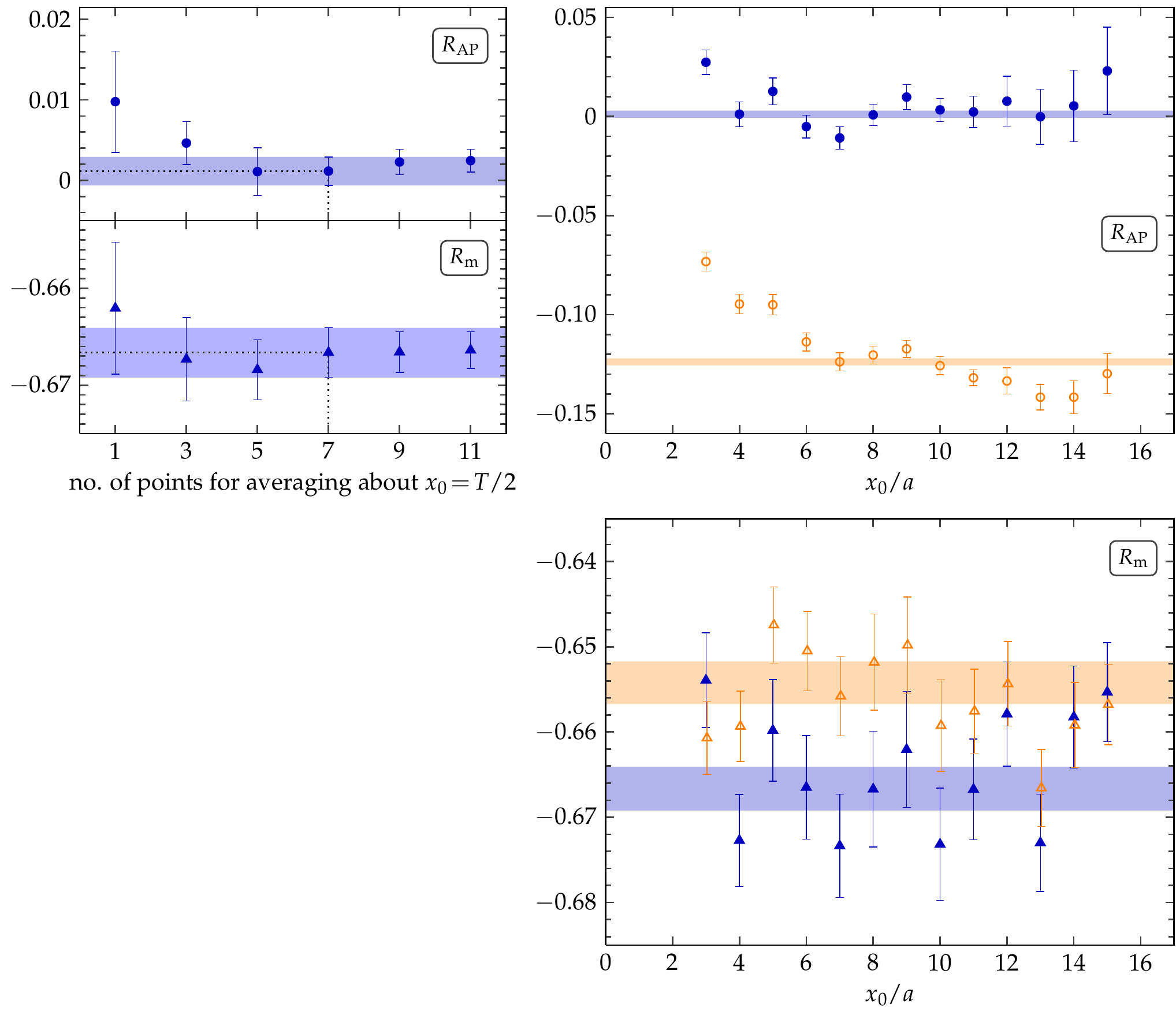}
\vspace{-0.875cm}
\caption{%
A typical data sample, showing $\RAP$ and $\Rm$ data at $L/a=12$ and
$\beta=6.3158$ for set~1.
\emph{Left:} Dependence on the plateau size used for the averaging, 
with the preferred choice indicated by dotted lines. 
\emph{Right:} Timeslice dependence of $R_X(x_0)$, if no plateau averaging
of the local PCAC masses is applied. 
Open symbols denote results using standard derivatives, while filled symbols
refer to those with improved lattice derivatives, eq.~\eqref{eqn:deriv}. 
For comparison, the shown (symmetric) error bands correspond to the results 
obtained from plateau-averaged masses.
}\label{fig:RX-plateaus}
}
%
For a careful error analysis we also looked at the dependence of the mean 
values and errors on the number of points entering the plateau average about 
$x_0=T/2$. 
This is exemplarily shown in the left plot of figure~\ref{fig:RX-plateaus}. 
There the leftmost point is the one at $x_0=T/2$ in the plots on the right. 
In the examples given here, the plateau average is approached fast. 
This also nicely reflects that a plateau average is less affected by local 
fluctuations, which could distort the mean value by more than its associated 
uncertainty.
If a plateau is visible, the averaging prescription --- being considered as
part of the definition of the $R_X$ as pointed out before --- leads to a 
more reliable estimate for the mean value and the corresponding error.

%
\FIGURE[t]{
\includegraphics[height=0.63\textheight]{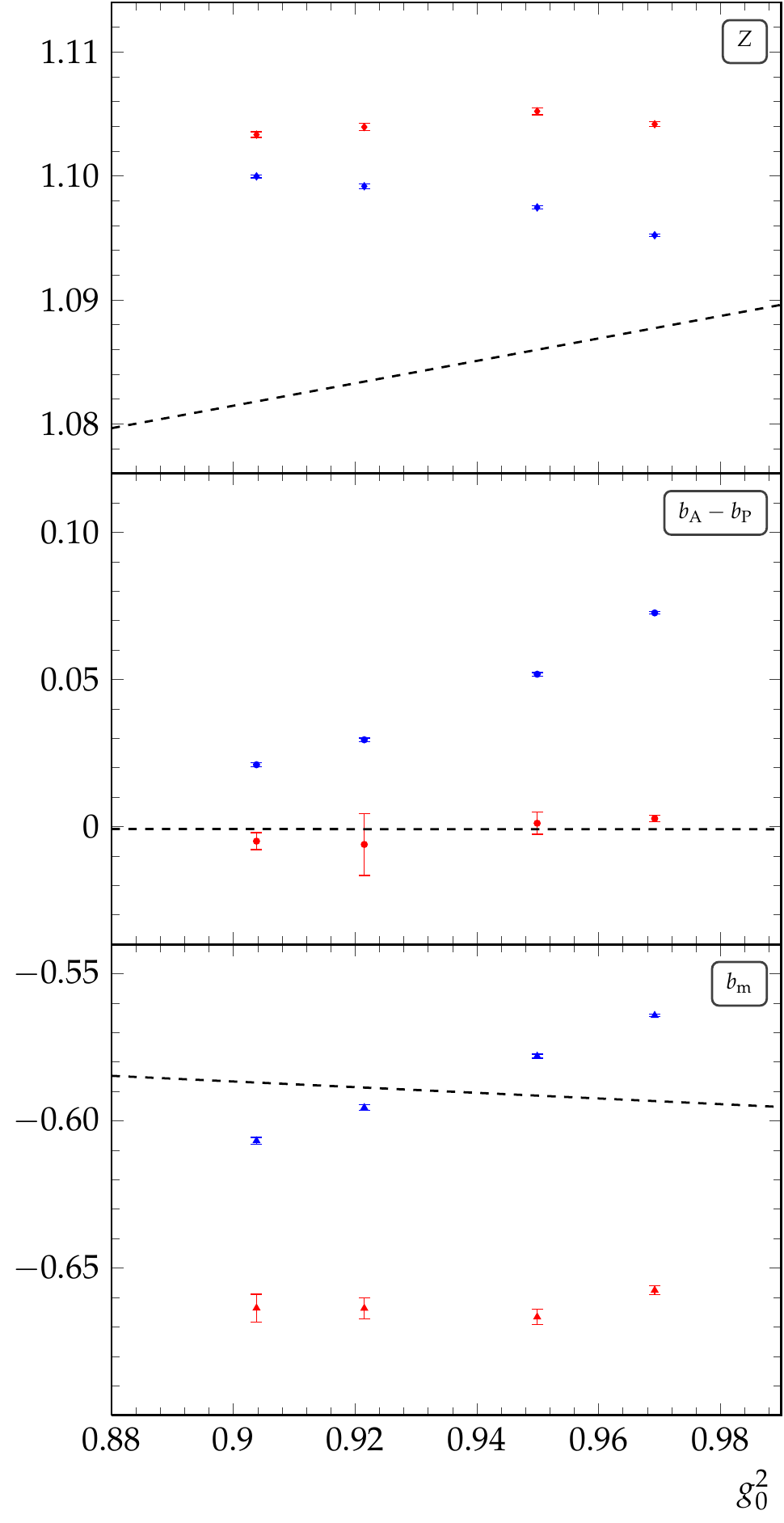}
\includegraphics[height=0.63\textheight]{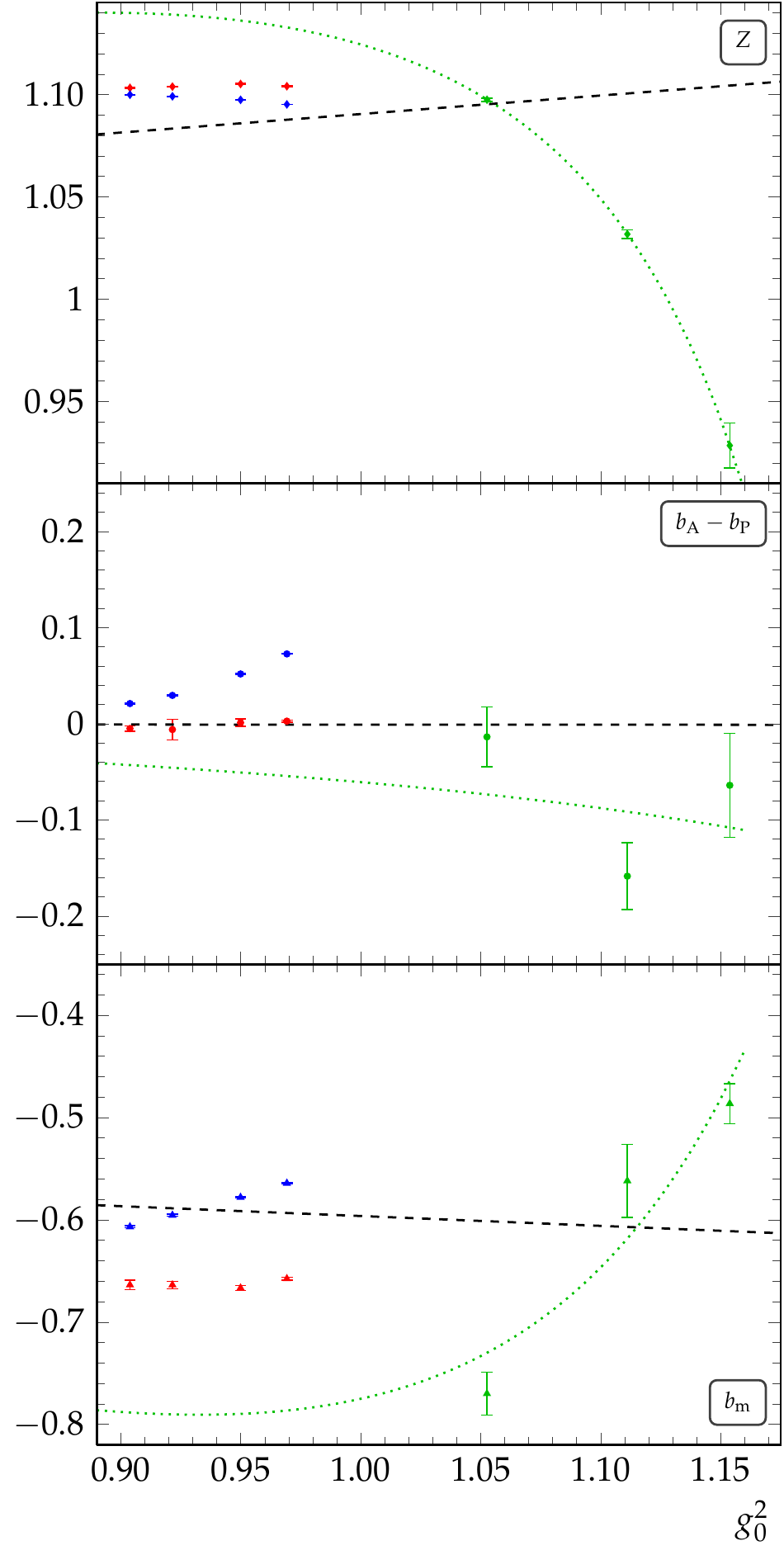}
\vspace{-0.375cm}
\caption{%
$g_0^2$-dependence of our final results for $Z$, $\ba-\bp$ and $\bm$ from 
top to bottom.
The dashed lines refer to the predictions from one-loop perturbation theory.
\emph{Left:} Results in the weak-coupling region of small-volume QCD
(constant physics condition 1) for set~1 (red points) and 
set~2 (blue points).
\emph{Right:} Same points together with the corresponding results of
subsection~\ref{Sec_res_RXcharm} in the region of larger couplings commonly 
employed in large-volume simulations (constant physics condition 2, 
green points).
}\label{fig:NP-results-RX-old-betas}
}
%
The dependence of our main estimates $R_X$ is visualized in the left plot of 
figure~\ref{fig:NP-results-RX-old-betas}. 
The non-perturbative data points show a smooth dependence on $g_0^2$ as 
expected, suggesting that the correct scaling of discretization errors 
proportional to $a$ for the improvement coefficients and $a^2$ for the 
renormalization constant is met. 
Red data points correspond to our preferred choice, set~1, whereas the blue 
points belong to set~2. 
The qualitative dependence on $g_0^2$ is comparable to the quenched 
case~\cite{HQET:pap2}. 
Note, however, that such a direct comparison at the quantitative level is 
of limited meaning only, since in the latter case the bare parameter 
region of the simulations referred to a smaller physical volume and the 
relation between the lattice spacing and the bare coupling is different from
the two-flavour theory.

In our fixed-volume computations, the lattice resolution gets smaller with 
decreasing $g_0^2$. Accordingly, the difference between both data sets 
decreases in this direction as it should be. Except for $\RAP$, set~1, the 
non-perturbative points show a partly 
significant deviation from the one-loop perturbative 
predictions \cite{impr:babp,impr:pap5} that are drawn as dashed lines in
figure~\ref{fig:NP-results-RX-old-betas}, too. 
For $\Rm$ this deviation is of $\Or(10\%)$ and for $R_Z$ still about $1\%$. 
One also observes the results of set~2 to have a larger curvature, but a 
smaller error in the range of bare couplings considered here.
This may be explained by the much larger valence quark mass, 
$Lm_{22}\,|_{\,\rm set2}\approx 5Lm_{22}\,|_{\,\rm set1}$, where the dynamics is 
governed by the heavy quark flavour and dominates the fluctuations induced
by the dynamical massless quarks in the gauge background.

The difference of $R_X$ between two improvement conditions 
is a measure for its $\Or(a^n)$ ambiguity.
In the extreme case of $\ba-\bp$, this difference in the results between
sets~1 and~2 is of the same order as the improvement coefficient itself so 
that in such a situation it appears rather tempting to just put the 
coefficient to zero. 
In the light of the qualitative discussion in 
subsection~\ref{Sec_impr_iccp}, however, it should be obvious that it is 
then not guaranteed that linear $a$-effects are absent after improvement.
Thus, an extrapolation to the continuum using an $\Or(a^2)$ model function 
for the cutoff effects might then give significantly wrong results. 
Based on this warning, let us now study such ambiguities quantitatively.
\subsubsection*{$\Or(a^n)$ ambiguities}
%
%
\FIGURE[t]{
\includegraphics[width=\textwidth]{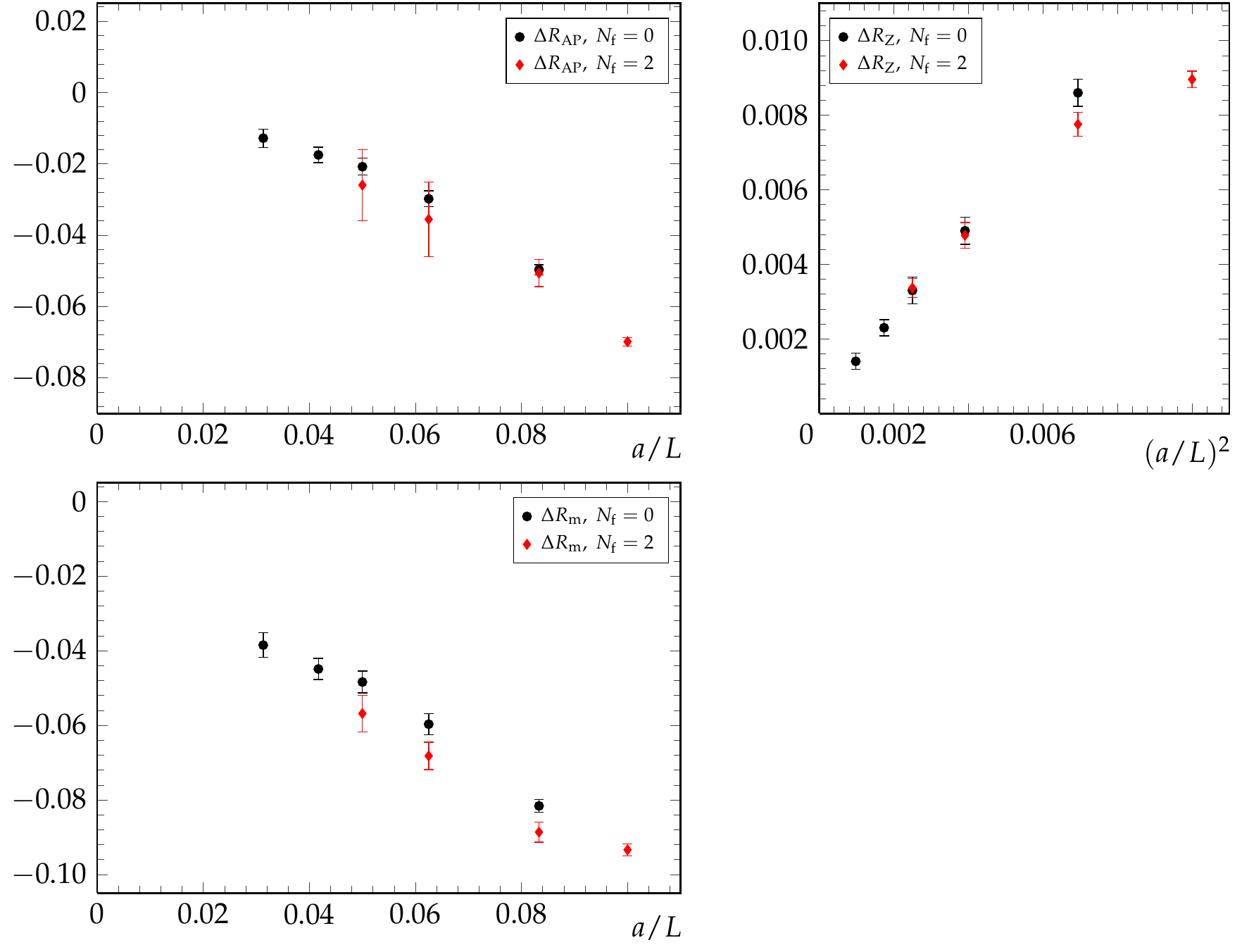}
\vspace{-0.875cm}
\caption{%
The differences $\Delta{R_X}$ of the estimators $R_X$ among the two sets
of results, as defined in eq.~\eqref{eqn:def-diff-set-1-2}.
Within errors, the differences approach zero with a rate proportional to
$a/L$ (for improvement coefficients) and $(a/L)^2$ (for the renormalization 
factor) as expected.
The earlier quenched results \cite{HQET:pap2} are added for comparison.
}\label{fig:bX_Z_diffs}
}
%
%
As a consequence of the very idea of the constant physics condition, any 
other estimate $R_{X}$ (i.e.~stemming from a different choice of 
improvement or renormalization condition) in general will yield a different 
functional dependence upon $g_0^2$ (and therefore $a/L$), but its 
differences are again smooth functions that must vanish in the continuum 
limit with a rate proportional to $a/L$ for improvement coefficients or 
$(a/L)^2$ for renormalization constants.
Regardless of their actual size, these intrinsic $\Or(a^n)$ ambiguities 
($n=1,2$) imply that rather than a numerical value at some given $\beta$, 
the important information lies in the correct \emph{$g_0^2$-dependence} of 
the estimators $R_{X}$, ${X}\in\{{\rm AP},{\rm m},Z\}$, obtained by working 
at \emph{constant physics} while varying $\beta$.
To demonstrate this, we investigated a few alternative improvement 
conditions, which are either already provided by the two quark mass choices
themselves, eqs.~\eqref{eqn:set1} and \eqref{eqn:set2}, or by defining the 
estimators $R_{X}$ with standard instead of improved derivatives.

Regarding the latter, we mentioned before that cutoff effects are visible 
between the use of standard and improved time derivatives, which are used 
to build the PCAC masses, and the reader may have wondered already about the 
large differences between these two choices of lattice derivatives in 
figure~\ref{fig:RX-plateaus}. 
But one should keep in mind that these results refer to an intermediate 
lattice spacing with $L/a=12$. 
When performing the full analysis for each data set with the standard 
derivatives, however, the differences to the results from employing improved 
derivatives were found to vanish in the limit $a\to 0$ as expected.

In addition, the two choices \eqref{eqn:set1} and \eqref{eqn:set2} made for
the non-vanishing valence quark mass themselves correspond to different 
improvement conditions to extract the estimators $R_X$ and so may yield an 
impression of the intrinsic cutoff ambiguities.
To this end we define  
\begin{align}
\Delta{R_X}(g_0) 
&\equiv 
\left.R_X(g_0)\,\right|_{\,\text{set1}}
-\left.R_X(g_0)\,\right|_{\,\text{set2}} \myt \,, &
X 
&\in   
\{{\rm AP},{\rm m},Z\} \myt \,.
\label{eqn:def-diff-set-1-2}
\end{align}
From eqs.~\eqref{eqn:estim-RAP} and~\eqref{eqn:estim-Rm} one expects a linear 
$a$-dependence and according to~\eqref{eqn:estim-RZ}, the cutoff dependence of 
$R_Z$ should be of $\Or(a^2)$.
For the values listed in table~\ref{tab:RX-direct-results} we show the 
results for $\Delta{R_X}$ as red points (diamonds) in 
figure~\ref{fig:bX_Z_diffs}. 
While the difference $\Delta{R_Z}$ in the right panel clearly exhibits a 
linear approach towards zero, the $\Or(a)$ ambiguities of $\ba-\bp$ and 
$\bm$ reflected in the left panel are quite small as well, and their 
magnitudes rapidly decrease as $a/L\rightarrow 0$.
The additional black points (circles) obtained in the quenched 
case~\cite{HQET:pap2} lie in the same ballpark, which can be attributed to 
the fact that the impact of the dynamical light quarks largely cancels out 
in the differences~\eqref{eqn:def-diff-set-1-2}; on top of that, the cutoff
ambiguities are dominated by the heavy quark mass content, which is about 
the same in both studies.
In summary, the scaling behaviour observed in the various differences of 
estimators \eqref{eqn:def-diff-set-1-2} confirms the theoretical expectation 
of the universality of the continuum limit and supports the reliability of 
the applied non-per\-turbative method also in the case of $\nf=2$ dynamical 
quarks.

\FIGURE[t]{
{%
\hspace*{1cm}
\includegraphics[width=0.50\textwidth]{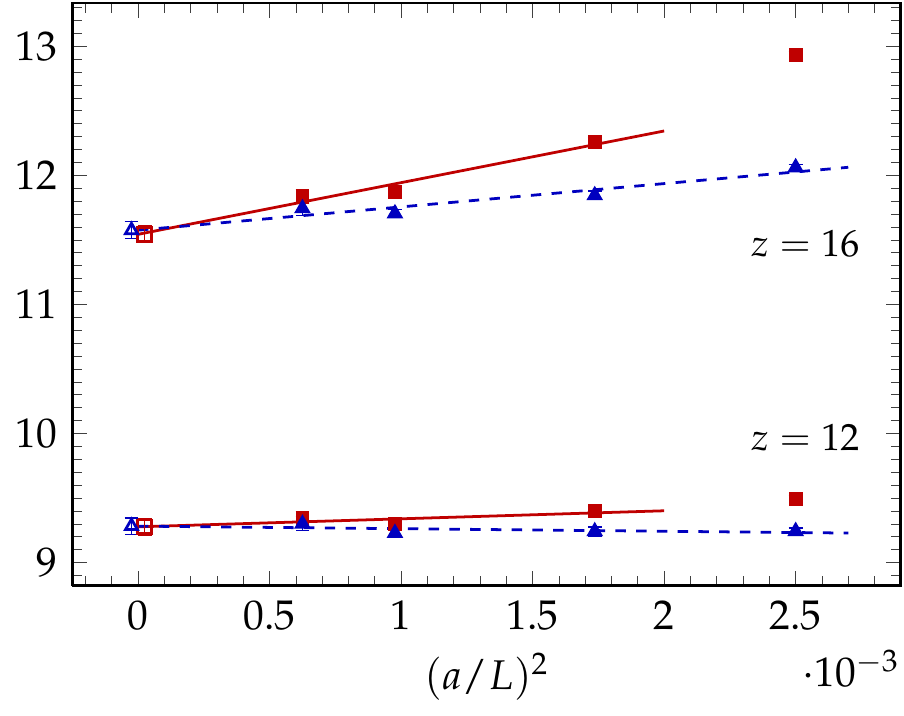}
\hspace*{1cm}
}%
\vspace{-0.25cm}
\caption{%
Unconstrained continuum extrapolations of $L_1\Gamps(z,g_0)$ for two fixed 
values of the renormalized quark, with different $\bm$ and $Z$ entering the 
latter as described in the text.
The data obtained via set~1 (red squares) and via set~2 (blue triangles) 
is fitted using three points (solid lines) and four points (dashed lines), 
respectively. 
}\label{fig:z12z16-test}
}
Although based on theoretical grounds discussed in 
subsection~\ref{Sec_impr_iccp} we expect that \emph{physical observables} 
involving sets of improvement coefficients from different constant physics 
conditions will have the same continuum limit, let us illustrate this in a 
concrete example. 
We consider the dimensionless combination $L_1\Gamps$, deriving 
from an effective energy that is extracted from the (improved) SF correlator 
of the time component of the axial~vector current in a finite volume $L_1^4$ with 
$L_1\approx0.5\,\fm$,
\begin{align}
L_1\Gamps(z,g_0) 
&=
\left.-L_1\sdrv0
\ln\left[\,\fa(x_0,M)\,\right]\,\right|_{\,x_0=L_1/2} \myt \,,
&
z&= L_1M \myt \,.
\label{gamps}
\end{align}
We confront its lattice spacing dependence using $\bm$ and $Z$ of set~1 with 
that using $\bm$ and $Z$ of set~2 to fix the dimensionless renormalization
group invariant (RGI) heavy quark mass to $z=12,16$ for lattice resolutions
of $L_1/a=20-40$. 
For further technical details, see section~\ref{Sec_lcps}.
Both data sets and its unconstrained continuum extrapolations linear in 
$(a/L)^2$ are displayed in figure~\ref{fig:z12z16-test} and extrapolate to 
the same continuum limit. Associated statistical errors are smaller or of the 
order of the symbol size.
The assumed quadratic scaling behaviour is clearly violated for the coarsest 
lattice at $z=16$ in set~1 though, because quark mass dependent cutoff
effects are substantial in that case.
Hence, we discarded the $L/a=20$ points in our fits of data set~1.
The fact that cutoff effects in $L_1\Gamps$ for both cases are larger for 
set~1 is not surprising, since the improvement condition of set~2 with 
$L_0m_{22}\approx 2.5$ and thus $L_1m_{22}\approx 5$ is much closer to the 
line in parameter space with $z=12,16$ along which $L_1\Gamps$ was computed.
This general behaviour present in our new data was also observed in the 
quenched case~\cite{HQET:pap2}.
The slope of the blue solid line at $z=16$ is approximately $6.5$ times larger 
than the corresponding one at $z=12$.
However, from the excellent agreement of the continuum limits in both cases 
one infers once more that our results correctly model the $g_0^2$-dependence
of $\bm$ and $Z$. 
No matter which set of improvement conditions is used, both entail 
convergence to the continuum limit with leading corrections of $\Or(a^2)$
in the $\Or(a)$ improved theory.
Consequently we can conclude that the ambiguity introduced by choosing 
either set~1 or set~2 vanishes in the continuum limit also for other values 
of $z$ and/or observables and universality holds.
\subsection{Results in the coupling region of large-volume simulations}
\label{Sec_res_RXcharm}
Besides in the parameter range discussed in the previous subsection, 
which is relevant, for instance, in the context of the matching calculations 
of HQET and QCD and to which we will come back again in the next subsection, 
the estimators $R_X$, $X\in\{{\rm AP},{\rm m},{Z}\}$, are also of strong 
interest in the parameter region, where lattice computations of hadron
masses and matrix elements in physically large volumes usually take place.
This is particularly the case for mesons involving the charm quark, because
large-volume studies in the quenched 
approximation \cite{mcbar:RS02,fds:final} have already given clear evidence 
for possibly substantial quark mass dependent cutoff effects down to
$a\approx 0.05\,\Fm$ so that a non-perturbative removal of the associated 
$\Or(a)$ corrections is highly desirable.
Thus, we here provide $\ba-\bp$, $\bm$ and $Z$ as functions in $g_0^2$ 
covering the range $5.2\le\beta\le 5.7$, which then will allows to 
interpolate to the typical $\beta$-values relevant for large-volume 
computations in two-flavour lattice QCD.

%
\TABLE[t]{
\centering
\small
\renewcommand{\arraystretch}{1.25}
\begin{tabular}{cccccccc} 
\toprule
 $L/a$ & $T/a$ & $\beta$ & $\hop_1$       & $Lm_{11}   $      & $\hop_2$       &  $Lm_{22}   $         & $\text{total}$ \\\midrule
  $12$ & $18$  & $5.20$  & $0.135500$     & $0.2495(30)$     & $0.134677$     &  $0.5006(28)$        & $4000        $ \\
       &       &         & $0.135700$     & $0.1584(32)$     & $0.134601$     &  $0.4956(28)$        & $4000        $ \\
       &       &         & $0.135800$     & $0.1051(30)$     & $0.134512$     &  $0.5015(23)$        & $4000        $ \\
       &       &         & $\hopc$        & $\it 0         $ & $\it   $       &  $\it 0.4986(36)$    &                \\\cmidrule(lr){1-8}
  $16$ & $24$  & $5.40$  & $\it 0.136645$ & $\it 0.0094(26)$ & $\it 0.135547$ &  $\it 0.4996(20)$    & $750\hpo$      \\\cmidrule(lr){1-8}
  $24$ & $36$  & $5.70$  & $\it 0.136704$ & $\it 0.0141(10)$ & $\it 0.136000$ &  $\it 0.5082(9)\hpo$ & $656\hpo$      \\
\bottomrule
\end{tabular}
\caption{%
Simulation parameters to compute $\bm$ and $\ba-\bp$ and $Z$ for a range
of bare couplings typically covered in large-volume simulations of $\nf=2$ 
lattice QCD.
The rightmost column gives the number of decorrelated gauge field 
configurations used.
The kinematical SF parameters are $(T/L,\theta)\!=\!(3/2,0)$, 
cf.~eqs.~\eqref{eqn:cond-coupl-5.5} and \eqref{eqn:kin-params-SF-bX-Z}.
}\label{tab:CharmP-parameters}

}
%
In table~\ref{tab:CharmP-parameters} we summarize the corresponding 
simulation parameters, where the relation between $\Lstar/a$ and $g_0$ from
\Ref{lat07:rainer} was taken to translate the constant physics condition~2,
eq.~\eqref{eqn:cond-coupl-5.5}, to pairs of $(L/a,\beta)$.
Owing to algorithmic instabilities caused by the appearance of very small,
unphysical eigenvalues of the SF Wilson-Dirac 
operator~\cite{Nf2SF:cutoff,ddhmc:stability}, the lattice with the coarsest 
lattice spacing (i.e.~$L/a=12$ at $\beta=5.2$) could only be simulated down 
to bare PCAC sea quark masses of about 
$a\ml\equiv am^{({\rm sea})}\approx 0.01$.
Therefore, our computations for this parameter set were based on dynamical
configurations at three sea quark mass values $\gtrsim 0.01$, made
available to us through the determination of $\za$ reported 
in~\cite{impr:za_nf2_2}, in order to extrapolate the results for $\beta=5.2$
to the chiral limit.

To eventually extract the estimators $R_X$ through 
eq.~\eqref{eqn:cond-coupl-5.5}, we here considered the 
setting~\eqref{eqn:set1} for the non-degenerate valence quarks so that the 
heavier PCAC quark mass $Lm_{22}$ had to be tuned to $\approx 0.5$, while 
the light one was again identified with the sea quark mass, 
$Lm_{11}=L\ml$, by setting $\kappa_1=\kapl$ ($=\kappa_c$ in case of 
$\beta=5.4,5.7$ only).
$Lm_{11}$ and $Lm_{22}$ are also listed in table \ref{tab:CharmP-parameters}.

%
\TABLE[t]{
\centering
\small
\renewcommand{\arraystretch}{1.25}
\begin{tabular}{CCCCCCC} 
\toprule
         &T/3\text{ avr'd}& x_0=T/2  &T/3\text{ avr'd}& x_0=T/2  &T/3\text{ avr'd}& x_0=T/2  \\\midrule
\hop_1   &  \MCC{0.135800}           &  \MCC{0.135700}           &  \MCC{0.135500}           \\
Lm_{11}  &  0.1051(30) &  0.1072(47) &  0.1584(32) &  0.1668(47) &  0.2495(30) &  0.2547(43) \\
Lm_{22}  &  0.5015(23) &  0.5015(37) &  0.4956(28) &  0.5016(39) &  0.5006(28) &  0.5041(37) \\\cmidrule(lr){1-7}
\ba-\bp  & -0.098(50)\,& -0.078(95)\;& -0.040(42)\;& -0.048(74)\;& -0.086(25)\;& -0.108(45)\;\\
\bm      & -0.453(28)\,& -0.413(43)\;& -0.513(25)\;& -0.511(43)\;& -0.485(20)\;& -0.424(36)\;\\
Z        &  0.9262(23) &  0.9194(38) &  0.9305(21) &  0.9225(35) &  0.9191(16) &  0.9058(31) \\
\bottomrule
\end{tabular}
\caption{%
Results at $\beta=5.2$ before extrapolating to the massless limit, stemming 
from averages over the central $T/3$ timeslices and from using $x_0=T/2$ 
only.
}\label{tab:CharmP-L12results}

}
%
For the $\beta=5.2$ data we do not observe a pronounced plateau in 
$R_X(x_0)$, and we thus gather in table~\ref{tab:CharmP-L12results} results 
stemming from our standard definition with an average taken over timeslices 
of the central third of the time extent, together with the corresponding 
results obtained at $x_0=T/2$.
Note that the improvement coefficients are well compatible within errors,
whereas for the renormalization constant this is only the case for the 
smallest light quark mass.
We hence decided to not alter our procedure of estimating these quantities 
using timeslice averages.
As for the chiral extrapolation $Lm_{11}=L\ml\to 0$ of the $\beta=5.2$ 
results, we investigated fits of $R_X$ to a constant as well as linear and 
quadratic fits using two or all three (light valence $=$ sea) quark mass 
points, which gave overall consistent results, since the dependence on the 
light quark mass is rather mild.
Two- and three-point weighted averages turned out to be most conservative
and as estimates in the massless limit we just quote the weighted averages 
of the numbers at the two lightest quark masses.
To account for a systematic error due to $\Or(a\ml)$ contaminations, 
we follow \Ref{impr:za_nf2_2} and inflate the error of the mean by its 
difference to the heaviest sea quark mass point linearly. 

\FIGURE[t]{
\includegraphics[width=\textwidth]{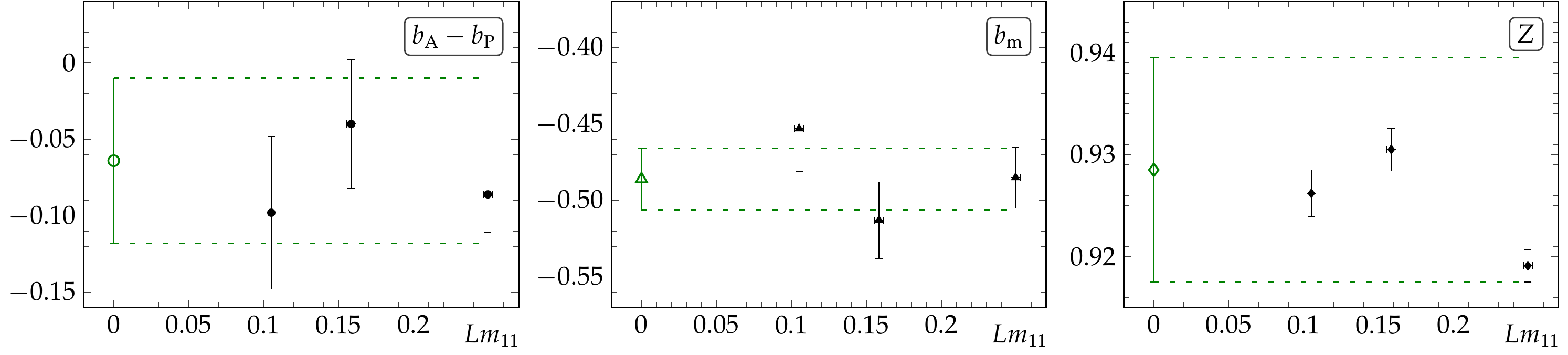}
\vspace{-0.75cm}
\caption{%
Extrapolations of the results at $\beta=5.2$ 
in table~\ref{tab:CharmP-L12results} to the chiral limit, $Lm_{11}=L\ml=0$. 
The weighted two-point averages shown in the plot include our estimate of the
$\Or(a\ml)$ systematic uncertainty as explained in the text and quoted in
table~\ref{tab:CharmP-results}.
}\label{fig:chiral-extrap-L0m11}
}
%
Our final results for all values of $\beta$ are summarized in 
table~\ref{tab:CharmP-results}.
Since simulations with massless dynamical quarks in this parameter region 
are numerically much more expensive, the $\beta=5.4$ and $5.7$ data samples 
have a much smaller statistics compared to the ones reported in 
subsection~\ref{Sec_res_RXweak}. 
Consequently, the estimates in this parameter region have an uncertainty,
which in general is one order of magnitude larger than in our small volume 
simulations at weaker couplings.
Even though one might generically expect a universal curve connecting the
results in distinct parameter regions, one encounters significantly different cutoff 
effects owing to different improvement conditions applied, as reflected on the
r.h.s of figure~\ref{fig:NP-results-RX-old-betas}.
%
\TABLE[t]{
\centering
\small
\renewcommand{\arraystretch}{1.25}
\begin{tabular}{ccccccc} 
\toprule
  $L/a$ &  $\beta$ & $\hop_1$       & $\ba-\bp           $ & $\bm              $ & $ Z                   $ \\ \midrule
   $12$ &  $5.20$  & $0.135500$     & $-0.086(25)\;      $ & $-0.485(20)\;     $ & $0.9191(16)           $ \\ 
        &          & $0.135700$     & $-0.040(42)\;      $ & $-0.513(25)\;     $ & $0.9305(21)           $ \\ 
        &          & $0.135800$     & $-0.098(50)\,      $ & $-0.453(28)\,     $ & $0.9262(23)           $ \\ 
        &          & $\hopc$        & $\it -0.064(32)(22)$ & $\it -0.486(19)(1)$ & $\it 0.9285(16)(94)$ \\\cmidrule(lr){1-6}
   $16$ &  $5.40$  & $\it 0.136645$ & $\it -0.159(37)    $ & $\it -0.562(36)   $ & $\it 1.0319(22)           $ \\\cmidrule(lr){1-6}
   $24$ &  $5.70$  & $\it 0.136704$ & $\it -0.014(21)    $ & $\it -0.770(21)   $ & $\it 1.0976(8)\hpo        $ \\
\bottomrule
\end{tabular}
\caption{%
Results of $\ba-\bp$, $\bm$ and $Z$ in the bare parameter region 
$5.2\le\beta\le 5.7$.
The quoted error on the results for $\beta=5.2$ is split into a statistical
and a systematic one, the latter accounting for possible $\Or(a\ml)$ 
contaminations from the extrapolation to the massless limit.
}\label{tab:CharmP-results}

}
%

For later use of our results it seems appropriate to represent them as
smooth functions in the interval $5.2\lesssim\beta\lesssim 5.7$.
As in the quenched computation~\cite{impr:babp}, we can readily parametrize 
$Z$, $\ba-\bp$ and $\bm$ by means of {Pad\'e} fits in $g_0^2$ 
that are constrained to pass asymptotically into the perturbative one-loop
predictions in the limit $g_0^2\to 0$.
In case of the renormalization factor $Z$, the fit
\begin{align}
Z\left(g_0^2\right) 
&= 
\left(1+0.090514\,g_0^2\right)\times 
\dfrac{1-0.3922\,g_0^4-0.2145\,g_0^6}{1-0.6186\,g_0^4}
\label{eqn:RZ-Pade-fit}
\end{align}
is found to describe our numbers very well, and the error to be attributed 
to this formula varies from about 1.2\% at $\beta=5.2$ to less than 0.1\% 
at $\beta=5.7$.
Due to the scatter in the results for $\ba-\bp$, we quote as {Pad\'e}
fit parametrization
\begin{align}
\left[\,\ba-\bp\,\right]\left(g_0^2\right) 
&= 
- 0.00093\,g_0^2\times 
\dfrac{1+23.59\,g_0^2}{1-0.6235\,g_0^2} \myt \,,
\label{eqn:RAP-Pade-fit}
\end{align}
which represents the data within two sigma of their overall uncertainty.
Note that accidentally the parameter in the numerator is close to the
quenched value, $23.306$, and yet an unconstrained linear fit would 
represent the data points nearly equally well. 
Finally, for the improvement coefficient $\bm$ we quote the interpolating 
formula
\begin{align} 
\bm\left(g_0^2\right) 
&= 
\left(-0.5-0.09623\,g_0^2\right)\times 
\dfrac{1-0.3737\,g_0^{10}}{1-0.5181\,g_0^4} \myt \,,
\label{eqn:Rm-Pade-fit}
\end{align}
still respecting the leading perturbative asymptotics.
When using this formula at $g_0^2$-values different from 
$\beta=5.2,5.4,5.7$, one should attribute to $\bm$ an uncertainty of the 
order of the error quoted in table~\ref{tab:CharmP-results} for the result at 
the nearest $\beta$.

\section{Lines of constant physics in finite-volume QCD}
\label{Sec_lcps}
As an immediate application of the results obtained in
subsection~\ref{Sec_res_RXweak}, we address the computation of the dependence
of heavy-light meson observables on the heavy quark mass in the continuum limit
of finite-volume QCD.  This is one of the key elements in the basic strategy
\cite{HQET:pap1} to non-perturbatively determine HQET parameters by a matching
to QCD.  Here, we only restrict the discussion to the fixing of the RGI heavy
quark mass in a given finite volume to a series of values covering the b-quark
mass down to the onset of the charm quark region, because it is this part,
where $\bm$ and $Z$ directly enter.  A preliminary account on the subsequent
computation of the heavy quark mass dependence of finite-volume QCD observables
was given in the report \cite{lat08:hqettests}.

Since we are eventually interested in quark mass dependent quantities evaluated
in the continuum limit of finite-volume QCD, we want to calculate them for a
series of bare parameters $(L/a,\beta,\kapl)$ such that the renormalized
parameters in the light quark sector are fixed and in this way physics is kept
constant along the approach to the continuum limit.  As before, $\kapl$ is the
hopping parameter of the mass degenerate dynamical light quarks, while their
PCAC mass is denoted by $\ml$.  In order to also fix the renormalized masses of
the non-degenerate valence quarks in units of a physical scale later on, one
first has to specify a suitable value of the renormalized SF coupling,
$\gbsq(L)$, which then, by its very definition \cite{SF:LNWW}, automatically
fixes the spatial extent of the SF cylinder as the only scale of our system.
Thus, in addition to $\theta=0.5$ and $T=L$, we choose
\begin{align}
\gbsq(L_0)    &= 2.989 \myt \,, &
L_0           &= \frac{L_1}{2} \myt \,, &
L_0\,\ml(L_0) &= 0
\label{eqn:cond-L1}
\end{align}
and thereby impose a {\em constant physics condition} on the renormalized
coupling and the mass of the sea quark doublet in a finite volume of extent
$L_0$.  Implicitly, this condition defines the volume characterized by
$L_1=2L_0$, in which the non-perturbative matching between HQET and QCD will be
performed \cite{lat07:hqetNf2}.  Although an exact knowledge of $L_1$ in
physical units is not yet needed at this stage, one can infer from the known
running of the SF coupling for $\nf=2$, c.f.~\Refs{alpha:Nf2_2,lat07:rainer},
that $L_1=2L_0\approx 0.5\,\Fm$.  Obviously, eq.~\eqref{eqn:cond-L1} is
identical to the corresponding settings of the constant physics `condition~1',
eq.~\eqref{eqn:cond-coupl-2.989}, used for the determination of $\ba-\bP$,
$\bm$ and $Z$ themselves in subsection~\ref{Sec_res_RXweak} and, in retrospect,
explains this so far supposedly arbitrary choice.

%
\TABLE[t]{
\centering
\small
\renewcommand{\arraystretch}{1.25}
\begin{tabular}{ccccccc} 
\toprule
  $L/a$ & $ \beta$ & $\zp(g_0,L/a)$   & set &  $ \ba\!-\!\bp      $ & $ \bm              $ & $ Z            $  \\\midrule
  $10 $ & $6.1569$ & $0.6065(9)\hpo$  & $1$ &  $-0.0000(12)\hpo   $ & $-0.6633(12)\hpo   $ & $1.10443(17)   $  \\[-0.2em]
        &          &                  & $2$ &  $+0.07852(53)      $ & $-0.56196(38)      $ & $1.09488(13)   $  \\[0.2em]
  $12 $ & $6.2483$ & $0.5995(8)\hpo$  & $1$ &  $-0.0016(8)\hpo\hpo$ & $-0.6661(9)\hpo\hpo$ & $1.10475(12)   $  \\[-0.2em]
        &          &                  & $2$ &  $+0.06284(33)      $ & $-0.57145(29)      $ & $1.09632(8)\hpo$  \\[0.2em]
  $16 $ & $6.4574$ & $0.5941(10)$     & $1$ &  $-0.0050(17)\hpo   $ & $-0.6674(23)\hpo   $ & $1.10455(17)   $  \\[-0.2em]
        &          &                  & $2$ &  $+0.03567(51)      $ & $-0.59147(63)      $ & $1.09888(10)   $  \\[0.2em]
  $20 $ & $6.6380$ & $0.5949(12)$     & $1$ &  $-0.0045(28)\hpo   $ & $-0.6692(27)\hpo   $ & $1.10379(17)   $  \\[-0.2em]
        &          &                  & $2$ &  $+0.02150(58)      $ & $-0.60763(78)      $ & $1.10021(10)   $  \\
\bottomrule   
\end{tabular}
\caption{%
Summary of the improvement coefficients and renormalization constants
slightly extra-/interpolated to ensure a fixed SF coupling in a volume with 
$L=L_0\approx 0.25\,\Fm$.
The results at $L/a=20$ had not to be extrapolated. 
$\zp$ was determined independently following the non-perturbative 
renormalization procedure of~\cite{msbar:Nf2}.
}\label{tab:RX-interp-results}

}
%
The peculiar value $\gbsq(L_0)=2.989$ was fixed by a simulation at $L_0/a=20$,
$T=L_0$, in conjunction with tentative interpolations in $\beta=6/g_0^2$ for
given $L_0/a\le 16$ to this target, based on the known dependence of the SF
coupling and the current quark mass on the bare parameters $(\beta,\kappa)$
available from the data of~\Ref{alpha:Nf2_2}.\footnote{%
Using the known
$\beta$-function and our experience from the quenched calculation
\cite{HQET:mb1m}, we can estimate that an uncertainty of about $0.04$ in the
coupling would translate via $L_1$ into an uncertainty in the b-quark mass of
at most $0.5\%$.} %
The condition of zero light sea quark mass in
\eq{eqn:cond-L1} is met by setting $\kapl$ to the critical hopping parameter,
$\kapc$, estimated again on basis of published data \cite{alpha:Nf2_2}, whereby
a slight mismatch of $|L_0\ml(L_0)|<0.05$ of this condition is tolerable in
practice.  The triples $(L_0/a,\beta,\kapl)$ resulting from this procedure are
those of table~\ref{tab:tune-LmPCAC}, used in our study of improvement and
renormalization factors described above.  In order to avoid a non-negligible
systematic error from small violations of the condition (\ref{eqn:cond-L1}) on
results of any subsequent computations in $L_1^4$ though, the $\beta$-values
were checked and fine-tuned further by additional simulations, which at the end
led to the values in the second column of table~\ref{tab:RX-interp-results}.
The corresponding small shifts in $\beta$ affect our previous estimates of
$\ba-\bP$, $\bm$ and $Z$ only at a negligible level, because any deviation from
the line of constant physics just entails a small change of the $\Or(a^2)$
effects that we nevertheless account for by interpolations in $\beta$.  The
resulting, slightly moved values of $\ba-\bP$, $\bm$ and $Z$ can be found in
table~\ref{tab:RX-interp-results}, too.

For the intended study of finite-volume QCD observables \cite{lat08:hqettests}
we consider SF correlation functions composed of a non-degenerate heavy-light
valence quark doublet, where the light valence quark mass is again chosen to be
equal to the vanishing sea quark mass adopted in the generation of the
dynamical gauge configurations, i.e. it is set to $\kapl=\kapc$.  Computing the
heavy quark mass dependence of observables then amounts to evaluate the
correlators on these configurations in a volume $L_1^3\times T$, $T=L_1$, for a
sequence of renormalized heavy valence quark mass values in a region extending
from charm to beauty.  For this purpose, the latter have to be fixed as
precisely as possible for the same given pairs $(L_0/a,\beta)$ that already fix
the volume $L_1$ via enforcing constant physics at $L_0=L_1/2$ through
eq.~\eqref{eqn:cond-L1}.  Therefore, starting
from~\eqref{eqn:renormPCAC-vs-mq}, we exploit the $\Or(a)$ improved relation
between the subtracted bare heavy quark mass $\mqh$ and the corresponding RGI
mass, viz.
\begin{align}
M
&=
h(L_0)\,\zm(g_0,L_0/a)\,\left(1+\bm(g_0)\,a\mqh\right)\,\mqh
+\Or(a^2) \myt \,,
\label{eqn:M-mqtil}
\end{align}
where
\begin{align}
\zm(g_0,L_0/a)
&=
\frac{Z(g_0)\,\za(g_0)}{\zp(g_0,L_0/a)} \myt \,, &
a\mqh
& =
\frac{1}{2}\left(\frac{1}{\kaph}-\frac{1}{\kapc}\right) \myt \,,
\label{eqn:Zm-mqh}
\end{align}
and the axial current renormalization constant $\za$ in dependence of $g_0$
is non-perturbatively know for $\nf=2$ from ref.~\cite{impr:za_nf2_2}.
The scale dependent renormalization constant $\zp$ was non-perturbatively 
determined in the course of our simulations in $L=L_0$, following exactly 
the prescription in~\Ref{msbar:Nf2}.
Its values relevant here (modulo a slight extrapolation in $\beta$ as in 
the case of $\ba-\bP$, $\bm$ and $Z$) are listed in the third column of 
table~\ref{tab:RX-interp-results}.
In eq.~\eqref{eqn:M-mqtil}, there also appears the factor
\begin{align}
h(L_0)
&\equiv
\frac{M}{\mbar(\mu_0)}=1.521(14) \myt \,, &
\mu_0
&=
\frac{1}{L_0}=\frac{2}{L_1} \myt \,,
\label{eqn:h-L0}
\end{align}
which represents the universal, regularization independent ratio of the RGI 
heavy quark mass, $M$, to the running quark mass $\mbar$ in the SF scheme at 
the renormalization scale $\mu_0$. 
$h(L_0)$ was evaluated by a reanalysis of available data on the 
non-perturbative quark mass renormalization in two-flavour QCD, also
published in~\Ref{msbar:Nf2}.

Having all ingredients of eqs.~\eqref{eqn:M-mqtil} -- \eqref{eqn:h-L0} at 
hand, particularly the improvement coefficient $\bm$ and the renormalization 
constant $Z$ obtained in this work, these equations can now 
straightforwardly be solved for any specific value of the dimensionless RGI 
heavy quark mass in $L_1$,
\begin{align}
z
&=
L_1 M \myt \,,
\end{align}
to yield the desired hopping parameter associated with the mass of the 
heavy valence quark flavour,
\begin{align}
\kaph
&=
\kaph(z,g_0)=
\left\{\dfrac{1}{\kapc}-\dfrac{1}{\bm}\left(
1-\sqrt{1+z\,\dfrac{4\,\bm}{[\,L_1/a\,]\,h\,\zm}}\,
\right)\right\}^{-1} \myt \,,
\label{eqn:kappah-z-repeated}
\end{align}
for which the heavy-light SF correlation functions and meson observables 
composed of them are to be evaluated.
For $\bm(g_0)$ and $Z(g_0)$ we employ the numbers belonging to 
the `set~1' (cf.~eq.~\eqref{eqn:set1}) of our results, because firstly, 
the $g_0^2$-dependence is weaker for $\bm$ and secondly, with the setting 
$L_0m_{22}\approx 0.5$ in their determination (see eq.~\eqref{eqn:set1}) we 
rather meet the condition that in \Ref{impr:babp} was found advantageous 
from the point of view of perturbation theory\footnote{
We recall in this context that it was explicitly demonstrated in 
subsection~\ref{Sec_res_RXweak} that a typical observable such as the
effective heavy-light meson energy approaches the same continuum limit,
when computed at a fixed value of $z$ in which either $\bm$ and $Z$ from
`set~1' or from `set~2' enter.
}.
Owing to the sign and the order of magnitude of the non-perturbative values 
for $\bm$ in the $\beta$-range relevant here, 
eq.~\eqref{eqn:kappah-z-repeated} has no real solutions for arbitrarily 
high $z$-values.
This demands to balance the upper bound on $z$ against the lattice 
resolutions one wants to simulate.
In our case of $L_1/a=2L_0/a=20,24,32,40$ we thus opted for nine values,
\begin{align}
z 
&\in 
\{4,6,7,9,11,13,15,18,21\} \myt \,,
\label{eqn:fixed-z-values}
\end{align}
which fairly span a range from around the charm to beyond the b-quark mass 
region.
Equation~\eqref{eqn:kappah-z-repeated} then yields the corresponding sets of
heavy-flavour hopping parameters for each $L_1/a$ collected in
table~\ref{tab:heavy-z-values-in-L1}, where solutions of 
\eqref{eqn:kappah-z-repeated} with $z=18,21$ only exist for inverse lattice
spacings of $L_1/a=24,32,40$.
%
\TABLE[t]{
\centering
\small
\renewcommand{\arraystretch}{1.25}
\begin{tabular}{CCCCCC}
\toprule
            &   z & \multicolumn{4}{C}{L/a}  \\\cmidrule(lr){3-6}
            &     &   20        &   24      &   32      &      40      \\\midrule
 \beta      &     &  6.1569     & 6.2483    & 6.4574    &  6.6380    \\
\kappa_{c} & \text{--} &     0.1360536	&     0.1359104	&     0.1355210	&     0.1351923	\\ 
\kappa_{1} &     4     &     0.1327094	&     0.1331966	&     0.1335537	&     0.1336432	\\ 
\kappa_{2} &     6     &     0.1309180	&     0.1317649	&     0.1325329	&     0.1328462	\\ 
\kappa_{3} &     7     &     0.1299824	&     0.1310257	&     0.1320117	&     0.1324413	\\ 
\kappa_{4} &     9     &     0.1280093	&     0.1294907	&     0.1309446	&     0.1316178	\\ 
\kappa_{5} &    11     &     0.1258524	&     0.1278628	&     0.1298401	&     0.1307738	\\ 
\kappa_{6} &    13     &     0.1234098	&     0.1261106	&     0.1286909	&     0.1299065	\\ 
\kappa_{7} &    15     &     0.1204339	&     0.1241815	&     0.1274876	&     0.1290126	\\ 
\kappa_{8} &    18     &     \text{---}	&     0.1206988	&     0.1255509	&     0.1276125	\\ 
\kappa_{9} &    21     &     \text{---}	&     0.1140810	&     0.1233865	&     0.1261232	\\ \bottomrule
\end{tabular}
\caption{%
Hopping parameters $\kappa_i\equiv\kaph^{(i)}$ from 
eq.~\eqref{eqn:kappah-z-repeated}, corresponding to fixed dimensionless RGI 
heavy quark masses $z=LM$ in a physical volume with $L=L_1\approx 0.5\,\Fm$ 
for different lattice resolutions. 
}\label{tab:heavy-z-values-in-L1}

}
%
\subsubsection*{Verification of $\Or(a)$ improvement}
\def\ZM{Z_{\rm M}}
For fixing $z$ to the values in (\ref{eqn:fixed-z-values}), one specific 
definition of the quark mass was applied, namely the relation between the 
bare subtracted heavy quark mass and the corresponding RGI 
mass~\eqref{eqn:M-mqtil}, where the previously determined improvement 
coefficient $\bm(g_0)$ and the renormalization constant $Z(g_0)$ enter. 
Nevertheless it appears tempting to have a look at equivalent definitions of 
the RGI heavy quark mass based on the PCAC relation, in order to provide 
another check of universality of the continuum limit in the $\Or(a)$ 
improved theory.
As a spin-off, this allows to also apply our non-perturbative knowledge of 
$[\ba-\bp](g_0)$ from subsection~\ref{Sec_res_RXweak} in an actual
calculation, even though the influence of the results on $\ba-\bp$ belonging 
to the `set~1' in this parameter region is expected to be numerically very
small.%
\footnote{%
Instead using the somewhat larger values of 'set 2' would not alter the outcome
of this paragraph because $\Delta{\RAP}$ vanishes in the continuum limit.
}%

For each $(L_1/a,\beta,\kapl)$, we take the heavy valence quark hopping 
parameters corresponding to the subset $z\in\{4,7,11,15,18,21\}$ and compute 
the bare PCAC masses $m_{ij}$ of eq.~\eqref{eqn:mpcac_x0-ij} from the 
heavy-light $(i\!=\!{\rm h}, j\!=\!{\rm l})$ and heavy-heavy 
$(i\!=\!j\!=\!{\rm h})$ axial current and pseudo-scalar density. 
With eqs.~\eqref{eqn:renorm-PCAC-relation-pm} -- \eqref{eqn:ren-off-diag-P} 
in mind, the renormalized PCAC quark masses then read
\begin{align}\label{eqn:renorm-PCAC-ij}
    \tfrac{1}{2}\,(\mri{i}+\mri{j})  &=
    \frac{\za(g_0)}{\zp(g_0,L_0/a)}\,
    \big\{1+[\ba-\bp](g_0)\,\half\left(a\mqi+a\mqj\right)\big\} m_{ij} 
    \myt \,.
\end{align}
Since we are interested in recovering the above values of $z=L_1M$ by a 
continuum extrapolation of suitable PCAC quark masses, we need to express 
eq.~\eqref{eqn:renorm-PCAC-ij} by RG invariants accordingly. 
Introducing the total renormalization factor 
$\ZM(g_0)=h(L_0)\za(g_0)/\zp(g_0,L_0/a)$, 
with $h(L_0)$ from eq.~\eqref{eqn:h-L0}, we end up with
\begin{align}\label{eqn:renorm-PCAC-z-ij}
    \tfrac{1}{2}(z_{i}+z_{j})  &=
    \ZM \big\{1+[\ba-\bp]\,\half\left(a\mqi+a\mqj\right)\big\} L_1m_{ij} 
    \myt \,. 
\end{align}
From the PCAC quark mass definition via the heavy-heavy current one thus 
directly has
%
\FIGURE[t]{
\includegraphics[width=\textwidth]{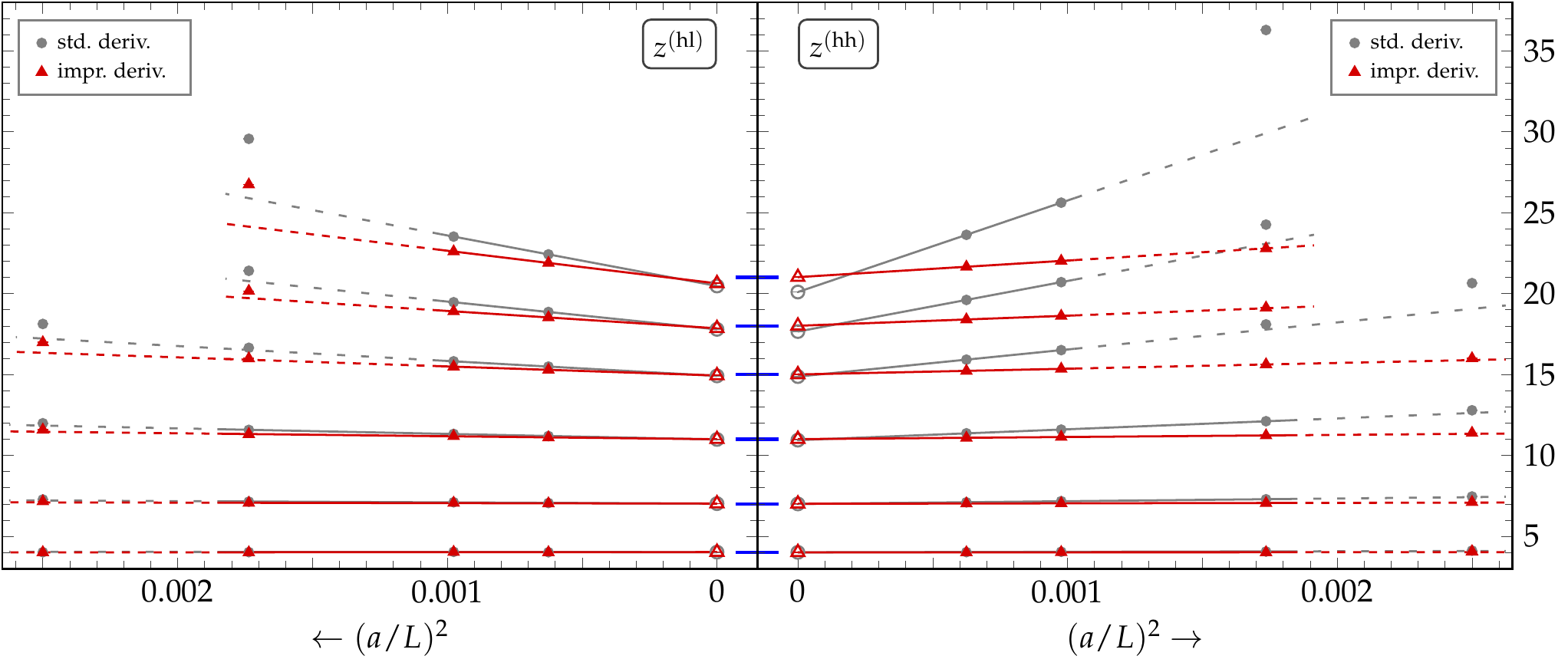}
\vspace{-0.875cm}
\caption{%
Continuum extrapolations of the dimensionless RGI PCAC heavy quark masses,
defined through either the heavy-light ({\it left}) or the heavy-heavy 
({\it right}) axial current, in a volume with $L=L_1$ for $\theta=0.5$. 
The statistical error is smaller than the symbol size. 
Dashed lines extend the fits to the points, which were not included in the
extrapolation, and the bold ticks on the $z$-axis indicate the subset of
$z$-values fixed via eq.~\eqref{eqn:M-mqtil} considered here.
}\label{fig:zhl-zhh}
}
%
\begin{align}\label{eqn:renorm-PCAC-z-hh}
    z^{({\rm hh})}  &=
    \ZM \big\{ 1 + [\ba-\bp]\,a\mqh \big\} L_1m_{\rm hh} \myt \,, 
\end{align}
while alternatively, $z$ can also by obtained through the quark mass 
definition via the heavy-light current as
\begin{align}\label{eqn:renorm-PCAC-z-hl-finite}
    z^{({\rm hl})}  &=
    2 \ZM \big\{ 
    1 + [\ba-\bp]\,\half\left(a\mqn{\rm h}+a\mql\right) \big\} L_1m_{\rm hl} 
    - z^{({\rm ll})} \myt \,; 
\end{align}
here, $z^{({\rm ll})}$ as in eq.~\eqref{eqn:renorm-PCAC-z-hh}, but with 
${\rm h}\to{\rm l}$. 
Since we have $\kappa_{\rm l}=\hopc$ for our light sea and valence quarks, 
this reduces to
\begin{align}\label{eqn:renorm-PCAC-z-hl}
    z^{({\rm hl})}  &=
    2 \ZM  \big\{ 1 + [\ba-\bp]\,\half a\mqn{\rm h} \big\} L_1m_{\rm hl} 
    \myt \,.
\end{align}

Based on universality arguments, we expect the continuum extrapolations of 
eqs.~\eqref{eqn:renorm-PCAC-z-hh} and~\eqref{eqn:renorm-PCAC-z-hl} to 
approach the corresponding values of $z$ fixed via the subtracted heavy 
quark mass, eq.~\eqref{eqn:M-mqtil}.
This is illustrated in figure~\ref{fig:zhl-zhh}. 
We show the results obtained from the heavy-light ($z^{({\rm hl})}$) and the
heavy-heavy ($z^{({\rm hh})}$) current and compare the estimators for the PCAC 
masses using standard and improved lattice derivatives as introduced 
earlier, c.f. eq.~\eqref{eqn:deriv}.
In qualitative agreement with our findings before, we in general observe 
that the cutoff effects increase with increasing values of $z$. 
The other expectation that cutoff effects are reduced when using improved 
lattice derivatives is clearly reflected as well. 
Whereas the continuum extrapolations for both lattice derivatives in the 
heavy-light case nicely match each other and coincide with the $z$-targets
even at large values of $z$, one observes a slight mismatch for $z\ge 15$ in 
the heavy-heavy current case.
This pattern can be explained by increasingly large quark mass dependent
$a$-effects inherent to the standard lattice derivative.  
Since we used improved lattice derivatives in the course of estimating 
$\ba-\bp$, $\bm$ and $Z$ underlying this analysis, it does not come as
a surprise that the smallest cutoff effects are seen for $z^{({\rm hh})}$ 
with improved derivatives. 
This is further supported by the fact that the extrapolating fit function
meets the data points even beyond the range included in the fit.
\subsubsection*{Error budget}
To summarize, $(L_1/a,\beta,\kapl,\kappa_i)$, 
with $\kappa_i\equiv\kaph^{(i)}$ and $i=1,\ldots,9$, finally 
constitute the parameters of the numerical valence quark propagator
calculations, from which the heavy-light correlation functions and the 
heavy quark mass dependence of the observables constructed from them are to
be extracted.
If for some reason (say, because of yet having used tentative results for
the quantities entering~\eqref{eqn:kappah-z-repeated}) a subsequent 
calculation was based on a marginally different set of $\kappa$-values,
the associated small mismatch in having fixed $z$ for each $L_1/a$ can 
easily be corrected by a small interpolation. 

At this point it still remains to quantify the uncertainty of $z$, which
comes into play through the various factors and coefficients it depends on
and which itself has to be propagated into any secondary quantity that is 
regarded as a function of $z$.
With the estimate $\Delta\za/\za\approx 0.36\%$ as quoted
in~\cite{impr:za_nf2_2} and $\Delta\zp/\zp$, $\Delta Z/Z$ as well as 
$\Delta \bm/\bm$ taken from table~\ref{tab:RX-interp-results}, ones obtains 
according to the standard rules of Gaussian error propagation an 
accumulated relative error on $z$ in the range 
$0.38\%\le(\Delta z/z)\le 0.41\%$. 
This holds for all values of $z$ in \eqref{eqn:fixed-z-values} and inverse
lattice resolutions $L_1/a$, but without the contribution from the 
continuum factor $h(L_0)$.
The latter represents with $\Delta h/h=0.92\%$ the dominating source of 
uncertainty in the total error budget of $\Delta z/z=1.01\%$, when the
fixing of $z$ proceeds as described here.

\section{Conclusions}
\label{Sec_concl}
In this paper we have non-perturbatively determined the renormalization 
constant and the improvement coefficients relating the renormalized current 
and subtracted quark mass in ${\rm O}(a)$ improved two-flavour lattice QCD.
To solve this rather technical and difficult numerical problem, improvement
and normalization conditions on lines of constant physics were imposed,
which ensures a smooth dependence of these quantities on the bare coupling
$g_0$ and the correct scaling of discretization errors proportional to $a$ 
(for the improvement coefficients) and $a^2$ (for the renormalization 
constant), respectively.  
In particular, our results open the way to a precision study of B-physics on 
the lattice, whose feasibility has already been demonstrated within the very 
useful test-laboratory of the quenched approximation over the last 
years~\cite{fb:roma2a,mbfb:Nf0,HQET:mb1m,HQET:param1m}.

An important element within these approaches is the separation of the scales
characterizing heavy-light systems and the introduction of small physical
volumes in order to perform QCD calculations with relativistic b-quarks. 
These simulations can only be performed by knowing the renormalization 
constants and improvement coefficients computed in the present paper for the 
dynamical two-flavour theory.
As explained in section~\ref{Sec_lcps}, with the help of the latter the 
renormalization group invariant heavy quark mass in finite-volume 
simulations of the ${\rm O}(a)$ improved theory can accurately be fixed to 
values from the b- down to the c-quark region.
Moreover, our results are also required when computing charmed observables
by making simulations on the same volumes used to study light flavour
physics. 

In the $\nf=2$ dynamical case, the programme of calculating the HQET 
parameters at $\Or(1/\mb)$ non-perturbatively by a matching to QCD, which
already employs part of the results of this work, is well advanced and 
progressing fast and will allow to determine several phenomenologically 
interesting quantities such as, for instance, the b-quark mass and B-meson 
decay constant.
As a further interesting direction for applications of the results of our
work let us mention the non-perturbative tests of the HQET expansion in the 
spirit of the quenched investigation~\cite{HQET:pap3}, which may greatly 
improve our confidence in the use of the effective theory.
This will be presented in a forthcoming publication.

\acknowledgments

We are indebted to G.~M.~de~Divitiis, M.~Della Morte, H.~Meyer, 
R.~Petronzio, H.~Simma, R.~Sommer and S.~Takeda for fruitful discussions and 
their contributions at an early stage of this work.
We also would like to thank R.~Sommer for a critical reading of the manuscript.
This work is part of the ALPHA Collaboration research programme.
We thank NIC/DESY and INFN for allocating computer time on the APE computers 
to this project as well as the staff of the computer centers at Zeuthen and
Rome for their support.
We acknowledge partial support by the European Community through 
EU Contract No.~MRTN-CT-2006-035482, ``FLAVIAnet''.
P.~F. and J.~H. also acknowledge financial support by the Deutsche 
Forschungsgemeinschaft (DFG) under grant \mbox{HE~4517/2-1}.
Some of the results presented here have already been published 
in~\cite{thesis:patrickf}.

\bibliography{lattice_ALPHA}

\providecommand{\href}[2]{#2}\begingroup\raggedright\begin{thebibliography}{10}

\bibitem{reviews:ChLFkaplan}
D.~B. Kaplan, {\it {Chiral Symmetry and Lattice Fermions}},
  \href{http://arXiv.org/abs/0912.2560}{{\tt 0912.2560}}.

\bibitem{impr:tmQCD}
R.~Frezzotti and G.~C. Rossi, {\it {Chirally improving Wilson fermions. I: O(a)
  improvement}},  {\em J. High Energy Phys.} {\bf 08} (2004) 007
  [\href{http://arXiv.org/abs/hep-lat/0306014}{{\tt hep-lat/0306014}}].

\bibitem{impr:sym1}
K.~Symanzik, {\it Continuum limit and improved action in lattice theories:
  1.~{Principles} and $\phi^4$ theory},  {\em Nucl. Phys.} {\bf B226} (1983)
  187.

\bibitem{impr:sym2}
K.~Symanzik, {\it Continuum limit and improved action in lattice theories:
  2.~{O}(${N}$) non-linear sigma model in perturbation theory},  {\em Nucl.
  Phys.} {\bf B226} (1983) 205.

\bibitem{impr:SW}
B.~Sheikholeslami and R.~Wohlert, {\it Improved continuum limit lattice action
  for {QCD} with {W}ilson fermions},  {\em Nucl. Phys.} {\bf B259} (1985) 572.

\bibitem{impr:onshell}
M.~{L\"uscher} and P.~Weisz, {\it On-shell improved lattice gauge theories},
  {\em Commun. Math. Phys.} {\bf 97} (1985) 59. Erratum: ibid. 98 (1985) 433.

\bibitem{impr:roma2_1}
G.~M. de~Divitiis and R.~Petronzio, {\it Non-perturbative renormalization
  constants on the lattice from flavor non-singlet ward identities},  {\em
  Phys. Lett.} {\bf B419} (1998) 311
  [\href{http://arXiv.org/abs/hep-lat/9710071}{{\tt hep-lat/9710071}}].

\bibitem{impr:babp}
{\bf ALPHA} Collaboration, M.~Guagnelli, R.~Petronzio, J.~Rolf, S.~Sint,
  R.~Sommer and U.~Wolff, {\it Non-perturbative results for the coefficients
  $b_{\rm m}$ and $b_{{\rm A}}-b_{{\rm P}}$ in {O}($a$) improved lattice
  {QCD}},  {\em Nucl. Phys.} {\bf B595} (2001) 44
  [\href{http://arXiv.org/abs/hep-lat/0009021}{{\tt hep-lat/0009021}}].

\bibitem{HQET:pap2}
{\bf ALPHA} Collaboration, J.~Heitger and J.~Wennekers, {\it Effective
  heavy-light meson energies in small-volume quenched {QCD}},  {\em J. High
  Energy Phys.} {\bf 02} (2004) 064
  [\href{http://arXiv.org/abs/hep-lat/0312016}{{\tt hep-lat/0312016}}].

\bibitem{SF:LNWW}
M.~{L\"uscher}, R.~Narayanan, P.~Weisz and U.~Wolff, {\it The {Schr\"odinger}
  functional: A renormalizable probe for non-abelian gauge theories},  {\em
  Nucl. Phys.} {\bf B384} (1992) 168
  [\href{http://arXiv.org/abs/hep-lat/9207009}{{\tt hep-lat/9207009}}].

\bibitem{SF:stefan1}
S.~Sint, {\it On the {Schr\"odinger} functional in {QCD}},  {\em Nucl. Phys.}
  {\bf B421} (1994) 135 [\href{http://arXiv.org/abs/hep-lat/9312079}{{\tt
  hep-lat/9312079}}].

\bibitem{SF:stefan2}
S.~Sint, {\it One-loop renormalization of the {QCD} {Schr\"odinger}
  functional},  {\em Nucl. Phys.} {\bf B451} (1995) 416
  [\href{http://arXiv.org/abs/hep-lat/9504005}{{\tt hep-lat/9504005}}].

\bibitem{impr:lett}
K.~Jansen {\em et.~al.}, {\it Non-perturbative renormalization of lattice {QCD}
  at all scales},  {\em Phys. Lett.} {\bf B372} (1996) 275
  [\href{http://arXiv.org/abs/hep-lat/9512009}{{\tt hep-lat/9512009}}].

\bibitem{HQET:pap1}
{\bf ALPHA} Collaboration, J.~Heitger and R.~Sommer, {\it Non-perturbative
  heavy quark effective theory},  {\em J. High Energy Phys.} {\bf 02} (2004)
  022 [\href{http://arXiv.org/abs/hep-lat/0310035}{{\tt hep-lat/0310035}}].

\bibitem{fb:roma2a}
M.~Guagnelli, F.~Palombi, R.~Petronzio and N.~Tantalo, {\it ${F}_{{\rm B}}$ and
  two scales problems in lattice {QCD}},  {\em Phys. Lett.} {\bf B546} (2002)
  237 [\href{http://arXiv.org/abs/hep-lat/0206023}{{\tt hep-lat/0206023}}].

\bibitem{mbfb:Nf0}
D.~Guazzini, R.~Sommer and N.~Tantalo, {\it Precision for {B}-meson matrix
  elements},  {\em J. High Energy Phys.} {\bf 01} (2008) 076
  [\href{http://arXiv.org/abs/0710.2229}{{\tt 0710.2229}}].

\bibitem{mcbar:RS02}
{\bf ALPHA} Collaboration, J.~Rolf and S.~Sint, {\it A precise determination of
  the charm quark's mass in quenched {QCD}},  {\em J. High Energy Phys.} {\bf
  12} (2002) 007 [\href{http://arXiv.org/abs/hep-ph/0209255}{{\tt
  hep-ph/0209255}}].

\bibitem{fds:final}
{\bf ALPHA} Collaboration, J.~Heitger and A.~{J\"uttner}, {\it Lattice cutoff
  effects for $f_{{\rm D}_{\rm s}}$ with improved {Wilson} fermions -- a final
  lesson from the quenched case},  {\em J. High Energy Phys.} {\bf 05} (2009)
  101 [\href{http://arXiv.org/abs/0812.2200}{{\tt 0812.2200}}].

\bibitem{lat08:mcharmNf2}
{\bf ALPHA} Collaboration, G.~{von Hippel}, R.~Sommer, J.~Heitger, S.~Schaefer
  and N.~Tantalo, {\it ${{\rm D}}_{{\rm s}}$ physics from fine lattices},  {\em
  PoS} {\bf LATTICE2008} (2008) 227 [\href{http://arXiv.org/abs/0810.0214}{{\tt
  0810.0214}}].

\bibitem{impr:ca_nf2}
{\bf ALPHA} Collaboration, M.~{Della Morte}, R.~Hoffmann and R.~Sommer, {\it
  Non-perturbative improvement of the axial current for dynamical {Wilson}
  fermions},  {\em J. High Energy Phys.} {\bf 03} (2005) 029
  [\href{http://arXiv.org/abs/hep-lat/0503003}{{\tt hep-lat/0503003}}].

\bibitem{impr:losalamos_3}
T.~Bhattacharya, R.~Gupta, W.-J. Lee, S.~R. Sharpe and J.~M.~S. Wu, {\it
  Improved bilinears in lattice {QCD} with non-degenerate quarks},  {\em Phys.
  Rev.} {\bf D73} (2006) 034504
  [\href{http://arXiv.org/abs/hep-lat/0511014}{{\tt hep-lat/0511014}}].

\bibitem{impr:losalamos_1}
T.~Bhattacharya, R.~Gupta, W.-J. Lee and S.~R. Sharpe, {\it {O}($a$) improved
  renormalization constants},  {\em Phys. Rev.} {\bf D63} (2001) 074505
  [\href{http://arXiv.org/abs/hep-lat/0009038}{{\tt hep-lat/0009038}}].

\bibitem{reviews:NPRrainer_nara}
R.~Sommer, {\it Non-perturbative {QCD}: renormalization, {O}($a$)-improvement
  and matching to {H}eavy {Q}uark {E}ffective {T}heory},  {\em Lectures given
  at {\it ILFTN Workshop on ``Perspectives in Lattice QCD''}, Nara, Japan, 31
  October -- 11 November 2005}
  [\href{http://arXiv.org/abs/hep-lat/0611020}{{\tt hep-lat/0611020}}].

\bibitem{impr:pap1}
M.~{L\"uscher}, S.~Sint, R.~Sommer and P.~Weisz, {\it Chiral symmetry and
  {O}($a$) improvement in lattice {QCD}},  {\em Nucl. Phys.} {\bf B478} (1996)
  365 [\href{http://arXiv.org/abs/hep-lat/9605038}{{\tt hep-lat/9605038}}].

\bibitem{impr:pap3}
{\bf ALPHA} Collaboration, M.~{L\"uscher}, S.~Sint, R.~Sommer, P.~Weisz and
  U.~Wolff, {\it Non-perturbative {O}($a$) improvement of lattice {QCD}},  {\em
  Nucl. Phys.} {\bf B491} (1997) 323
  [\href{http://arXiv.org/abs/hep-lat/9609035}{{\tt hep-lat/9609035}}].

\bibitem{impr:za_nf2_2}
{\bf ALPHA} Collaboration, M.~{Della Morte}, R.~Sommer and S.~Takeda, {\it On
  cutoff effects in lattice {QCD} from short to long distances},  {\em Phys.
  Lett.} {\bf B672} (2009) 407 [\href{http://arXiv.org/abs/0807.1120}{{\tt
  0807.1120}}].

\bibitem{lat07:hqetNf2}
{\bf ALPHA} Collaboration, M.~{Della Morte}, P.~Fritzsch, J.~Heitger, H.~B.
  {Meyer}, H.~Simma and R.~Sommer, {\it Towards a non-perturbative matching of
  {HQET} and {QCD} with dynamical light quarks},  {\em PoS} {\bf LAT2007}
  (2007) 246 [\href{http://arXiv.org/abs/0710.1188}{{\tt 0710.1188}}].

\bibitem{lat08:hqettests}
{\bf ALPHA} Collaboration, M.~{Della Morte}, P.~Fritzsch, J.~Heitger and
  R.~Sommer, {\it Non-perturbative quark mass dependence in the heavy-light
  sector of two-flavour {QCD}},  {\em PoS} {\bf LATTICE2008} (2008) 226
  [\href{http://arXiv.org/abs/0810.3166}{{\tt 0810.3166}}].

\bibitem{cern:Nf2light_1}
L.~{Del Debbio}, L.~Giusti, M.~{L\"uscher}, R.~Petronzio and N.~Tantalo, {\it
  {QCD} with light {W}ilson quarks on fine lattices {(I)}: first experiences
  and physics results},  {\em J. High Energy Phys.} {\bf 02} (2007) 056
  [\href{http://arXiv.org/abs/hep-lat/0610059}{{\tt hep-lat/0610059}}].

\bibitem{lat07:rainer}
{\bf ALPHA} Collaboration, M.~{Della Morte}, P.~Fritzsch, B.~Leder, H.~B.
  {Meyer}, H.~Simma, R.~Sommer, S.~Takeda, O.~Witzel and U.~Wolff, {\it
  Preparing for ${N}_{\rm f}=2$ simulations at small lattice spacings},  {\em
  PoS} {\bf LAT2007} (2007) 255 [\href{http://arXiv.org/abs/0710.1263}{{\tt
  0710.1263}}].

\bibitem{hmc:hasenb1}
M.~Hasenbusch, {\it Speeding up the {Hybrid-Monte-Carlo} algorithm for
  dynamical fermions},  {\em Phys. Lett.} {\bf B519} (2001) 177
  [\href{http://arXiv.org/abs/hep-lat/0107019}{{\tt hep-lat/0107019}}].

\bibitem{hmc:hasenb2}
M.~Hasenbusch and K.~Jansen, {\it Speeding up lattice {QCD} simulations with
  clover-improved {W}ilson fermions},  {\em Nucl. Phys.} {\bf B659} (2003) 299
  [\href{http://arXiv.org/abs/hep-lat/0211042}{{\tt hep-lat/0211042}}].

\bibitem{Nf2SF:algo}
{\bf ALPHA} Collaboration, M.~{Della Morte}, F.~Knechtli, J.~Rolf, R.~Sommer,
  I.~Wetzorke and U.~Wolff, {\it Simulating the {Schr\"odinger} functional with
  two pseudo-fermions},  {\em Comput. Phys. Commun.} {\bf 156} (2003) 62
  [\href{http://arXiv.org/abs/hep-lat/0307008}{{\tt hep-lat/0307008}}].

\bibitem{Nf2SF:autocorr}
{\bf ALPHA} Collaboration, H.~B. {Meyer}, H.~Simma, R.~Sommer, M.~{Della
  Morte}, O.~Witzel and U.~Wolff, {\it Exploring the {HMC} trajectory-length
  dependence of autocorrelation times in lattice {QCD}},  {\em Comput. Phys.
  Commun.} {\bf 176} (2007) 91
  [\href{http://arXiv.org/abs/hep-lat/0606004}{{\tt hep-lat/0606004}}].

\bibitem{MCerr:ulli}
{\bf ALPHA} Collaboration, U.~Wolff, {\it Monte {C}arlo errors with less
  errors},  {\em Comput. Phys. Commun.} {\bf 156} (2004) 143
  [\href{http://arXiv.org/abs/hep-lat/0306017}{{\tt hep-lat/0306017}}].

\bibitem{impr:pap5}
{\bf ALPHA} Collaboration, S.~Sint and P.~Weisz, {\it Further results on
  {O}($a$) improved lattice {QCD} to one-loop order of perturbation theory},
  {\em Nucl. Phys.} {\bf B502} (1997) 251
  [\href{http://arXiv.org/abs/hep-lat/9704001}{{\tt hep-lat/9704001}}].

\bibitem{Nf2SF:cutoff}
{\bf ALPHA} Collaboration, M.~{Della Morte}, R.~Hoffmann, F.~Knechtli and
  U.~Wolff, {\it {Impact of large cutoff-effects on algorithms for improved
  Wilson fermions}},  {\em Comput. Phys. Commun.} {\bf 165} (2005) 49
  [\href{http://arXiv.org/abs/hep-lat/0405017}{{\tt hep-lat/0405017}}].

\bibitem{ddhmc:stability}
L.~D. Debbio, L.~Giusti, M.~{L\"uscher}, R.~Petronzio and N.~Tantalo, {\it
  Stability of lattice {QCD} simulations and the thermodynamic limit},  {\em J.
  High Energy Phys.} {\bf 02} (2006) 011
  [\href{http://arXiv.org/abs/hep-lat/0512021}{{\tt hep-lat/0512021}}].

\bibitem{alpha:Nf2_2}
{\bf ALPHA} Collaboration, M.~{Della Morte}, R.~Frezzotti, J.~Heitger, J.~Rolf,
  R.~Sommer and U.~Wolff, {\it Computation of the strong coupling in {QCD} with
  two dynamical flavours},  {\em Nucl. Phys.} {\bf B713} (2005) 378
  [\href{http://arXiv.org/abs/hep-lat/0411025}{{\tt hep-lat/0411025}}].

\bibitem{msbar:Nf2}
{\bf ALPHA} Collaboration, M.~{Della Morte}, R.~Hoffmann, F.~Knechtli, J.~Rolf,
  R.~Sommer, I.~Wetzorke and U.~Wolff, {\it Non-perturbative quark mass
  renormalization in two-flavor {QCD}},  {\em Nucl. Phys.} {\bf B729} (2005)
  117 [\href{http://arXiv.org/abs/hep-lat/0507035}{{\tt hep-lat/0507035}}].

\bibitem{HQET:mb1m}
{\bf ALPHA} Collaboration, M.~{Della Morte}, N.~Garron, M.~Papinutto and
  R.~Sommer, {\it Heavy quark effective theory computation of the mass of the
  bottom quark},  {\em J. High Energy Phys.} {\bf 01} (2007) 007
  [\href{http://arXiv.org/abs/hep-ph/0609294}{{\tt hep-ph/0609294}}].

\bibitem{HQET:param1m}
{\bf ALPHA} Collaboration, B.~Blossier, M.~{Della Morte}, N.~Garron and
  R.~Sommer, {\it {HQET} at order $1/m$: {I.} {Non-perturbative} parameters in
  the quenched approximation},  \href{http://arXiv.org/abs/1001.4783}{{\tt
  1001.4783}}.

\bibitem{HQET:pap3}
{\bf ALPHA} Collaboration, J.~Heitger, A.~{J\"uttner}, R.~Sommer and
  J.~Wennekers, {\it Non-perturbative tests of heavy quark effective theory},
  {\em J. High Energy Phys.} {\bf 11} (2004) 048
  [\href{http://arXiv.org/abs/hep-ph/0407227}{{\tt hep-ph/0407227}}].

\bibitem{thesis:patrickf}
P.~Fritzsch, {\it {B}-meson properties from non-perturbative matching of {HQET}
  to finite-volume two-flavour {QCD}},  {\em {\rm Ph.D. Thesis}} (2009). {\rm
  {Universit\"at M\"unster}}.

\end{thebibliography}\endgroup
\bibliographystyle{JHEP-2}
\end{document}